\documentclass[aps,prd,english,notitlepage,nofootinbib]{revtex4-1}
\usepackage[T1]{fontenc}
\usepackage[latin9]{inputenc}
\usepackage[usenames,dvipsnames]{color}  
\definecolor{darkred}{rgb}{0.6,0,0}
\definecolor{brown}{rgb}{0.59, 0.29, 0.0}

\definecolor{darkcyan}{RGB}{0, 111, 111}
\definecolor{dgreen}{rgb}{0,0.5,0}
\usepackage[colorlinks=true,citecolor=darkred,urlcolor=darkred, pdfborder={0 0 0}]{hyperref}
\usepackage{ulem}
 
\newcommand {\ignore}[1]{}
\usepackage{amsmath,amssymb,mathtools,graphicx,fancyhdr}
\usepackage{multirow}
\usepackage{graphics, setspace}
\usepackage{subcaption}
\begin{document}

\title{Charged lepton-flavor violating constraints to non-unitarity in the Linear Seesaw scheme}
\author{Jes\'us Miguel Celestino-Ram\'irez~}
\email{jesus.celestino@cinvestav.mx}
\affiliation{Departamento de F\'{\i}sica, Centro de   Investigaci{\'o}n y de Estudios Avanzados del IPN\\ Apartado Postal   14-740 07000 Ciudad de M\'exico, Mexico}
\author{O. G. Miranda}
\email{omar.miranda@cinvestav.mx}
\affiliation{Departamento de F\'{\i}sica, Centro de   Investigaci{\'o}n y de Estudios Avanzados del IPN\\ Apartado Postal   14-740 07000 Ciudad de M\'exico, Mexico}

\begin{abstract}
We analyze the non-unitary effects in the linear seesaw mechanism using the current constraints and future sensitivity of the charged Lepton Flavor Violation (cLFV) processes. We perform a random scan confronting the non-unitary parameters with the limits to the cLFV processes. We show our results for the normal and inverted ordering of the oscillation data. We also discuss the equivalence of different parametrizations of the non-unitary mixing matrix and show our results in terms of the different parameterizations. We found that the stronger restrictions in the non-unitary parameter space come from rare muon decay searches.
\end{abstract}
\maketitle

\section{Introduction}
Just after the proposal of the Standard Model (SM), the efforts for a
new physics model that explains the observed nature of elementary
particles as minimal as possible have been constantly growing. The confirmation of neutrino oscillations, which implies that neutrinos
have a mass different from zero, boosted the search
for new physics in the neutrino sector. Of particular interest are the
seesaw
mechanisms~\cite{Minkowski:1977sc,Mohapatra:1979ia,Gell-Mann:1979vob,Yanagida:1979as,Schechter:1980gr,Foot:1988aq}
that aim precisely to explain the smallness of the neutrino masses
involved in oscillations, adding Majorana right-handed heavy neutral states. These heavy right-handed neutrinos are singlets
under the SM gauge symmetry and generate the suppression relation between the light and heavy neutrino mass. Moreover, low-scale seesaw mechanisms like the
inverse~\cite{Mohapatra:1986bd} and
linear~\cite{Akhmedov:1995vm,Malinsky:2005bi} seesaw have the extra
characteristic that they could be realized at a relatively lower mass scales, making them more attractive, as they can
be tested in the near future accelerators~\cite{Batra:2023mds}. Besides,
there are also searches for indirect evidence of the Heavy Neutral Leptons (HNL) due to the
non-unitary of the $3\times 3$ lepton mixing
matrix~\cite{Miranda:2020syh,Escrihuela:2015wra}.

Besides the different searches for signatures of neutral heavy leptons
in accelerators through direct production, it is also
expected that HNL may induce lepton flavor violation. This work
will focus on charged lepton flavor violation (cLFV) observables, studying
current constraints on the branching ratios and future expected
sensitivities. For cLFV observables, the predicted branching ratio will
depend on the neutral mass eigenstates and on the new mixing
matrix parameters. Given a neutrino mass pattern, we can compute the branching ratio in terms of the light mixing sector, allowing for constraints on the non-unitary parameters. We will study this phenomenology in the context of the linear seesaw mechanism. Previous works in a similar direction exist in the literature~\cite{Batra:2023mds,CentellesChulia:2024uzv}. For example, the case of the inverse seesaw was already studied~\cite{Garnica:2023ccx}, and thorough studies with previous data have also been carried out~\cite{Forero:2011pc}.

This work is organized as follows: section II
describes the general seesaw mechanism, the low-scale (linear)
mechanism and the nonunitary effects in both schemes. The
section III describes the flavor violation processes 
and contains all the
numerical aspects of our analysis and all the parametrizations that we
follow. Section IV presents the sensitivity of the nonunitary
parameters and the phase space of the linear seesaw in these flavor
violation processes and shows our conclusions.

\section{seesaw mechanism}
 A simple and natural way to explain
  the neutrino tiny masses is the well-known seesaw type-I mechanism \cite{Mohapatra:1979ia,Minkowski:1977sc, Yanagida:1979as, Gell-Mann:1979vob}. In this model, the mass matrix has the following form:  
\begin{equation}
\label{masses}
    M_{n\times n}=\begin{pmatrix}
0 & M_D \\ 
M^T_D & M_R 
\end{pmatrix} .
\end{equation}
There is no limit for the number of singlets that can be added in the theory. The physical neutrino masses are obtained by diagonalizing Eq.~(\ref{masses})
\begin{equation}
\label{diagonalize}
   U^T M_{n\times n}U= M_{diag} .
\end{equation}
There are different ways to compute the matrix U. Let's consider that, at leading order, it takes the form~\cite{Kanaya:1980cw,Schechter:1981cv}
\begin{equation}
\label{complete U}
U=\text{\~U} \cdot V,
\end{equation}
where \~U=$e^{iH}$, with
\begin{equation}H=\begin{pmatrix}0 & S \\ S^\dagger & 0 \end{pmatrix}, \quad 
V=\begin{pmatrix}
V_{1_{3\times 3}} & 0 \\ 
0 & V_{2_{m \times m}} 
\end{pmatrix} .
\end{equation}
We assume that the main contribution for diagonalizing the neutrinos mass matrix comes from $V_{1_{3\times 3}}$ and $V_{2_{m \times m}}$ while $S$ should give play a sub-leading role in this diagonalization process. After expanding in series we can get 
\begin{equation}
\text{\~U}\simeq\begin{pmatrix}
(\mathbb{I}-\frac{1}{2}SS^{\dagger})_{3\times 3} & iS_{3\times m} \\ 
iS^{\dagger}_{m \times 3} &(\mathbb{I}-\frac{1}{2}S^{\dagger}S)_{m\times m}
\end{pmatrix} ,
\end{equation}
with $m=n-3$.
Considering that we must fulfill Eq.~(\ref{diagonalize}) we can express \~U as~\cite{Forero:2011pc,Garnica:2023ccx} 
\begin{equation}
\label{rotation}
\text{\~U}= \begin{pmatrix}
\mathbb{I}_{3 \times 3}-\frac{1}{2}(M_D^*(M^*_R)^{-1}M^{-1}_RM^{T}_D)_{3\times 3} & (M^*_D(M_R^
*)^{-1})_{3\times m} \\ 
(M_R^{-1}M^T_D)_{m \times 3} &  \mathbb{I}_{3 \times 3}-\frac{1}{2}(M^{-1}_RM^{T}_DM_D^*(M^*_R)^{-1})_{m\times m}
\end{pmatrix}  .
\end{equation}
Using Eqs.~(\ref{diagonalize}) and (\ref{rotation}), we get: 
\begin{align}
    m_{diag}=&(V_1^Tm_{\nu}V_1)_{3\times 3}\\ 
    M^{diag}_{N}=& (V_2^TM_{N}V_2)_{m\times m}.
\end{align}
We can observe that $V_1$ and $V_2$ are unitary matrices and diagonalize the light and heavy sectors respectively. At leading order, $m_{\nu}$ and $m_{N}$ are given by:
\begin{align}
    m_{\nu}\simeq&-(M_DM^{-1}_RM^{T}_D)_{3\times 3} \\
    M_N\simeq&M_R
\end{align}

\subsection*{Nonunitarity effects}
The new extra heavy neutrino massive states induce effects in the light sector. We can observe that the light sector is no longer unitarity, unlike the standard oscillation paradigm. Nonunitary signals have been searched for in the literature since it will be a clear probe of physics beyond the Standard Model~\cite{Gronau:1984ct,Nardi:1994iv,Atre:2009rg,Escrihuela:2016ube,Fernandez-Martinez:2016lgt,Blennow:2023mqx,Forero:2021azc,Denton_2022,Dutta:2019hmb}. 
To study the nonunitary effects, we can use the rectangular matrix $K_{ij}$ that characterizes the charge current interaction for leptons. It can be seen in the following lagrangian:

\begin{equation}
    \mathcal{L}=-\frac{g}{\sqrt{2}}W^{-}_{\mu}\sum^{3}_{i=1}\sum^{n}_{j=1}K_{ij}\Bar{l}_i\gamma^{\mu}P_L \nu_j +h.c.,
\end{equation}
where 
\begin{equation}
 K_{ij}=\sum^{3}_{c=1}\Omega^*_{ci}U_{cj}.   
\end{equation}
and the subscript $j$ runs along all the neutrino states, including the heavy ones. 
The $\Omega_{ci}$ is the charged lepton mixing matrix; in this work, we assume it is diagonal ($\Omega_{ci}=\delta_{ci}$), so the matrix K depends totally on the neutrino mixing matrix.  The K matrix will contain 3(n-2) mixing angles and 3(n-2) phases, that is the necessary number of parameters to describe the physical rotations~(see for example~\cite{Giunti:2007ry,Rodejohann:2011vc,Valle:2015pba}). It is useful to describe the matrix K as block matrices: 
\begin{equation}
  \label{kappa}
    K=(N,S) ,
\end{equation}
where N describes the mixing with the light sector, and S with the heavy one. Using Eqs.~(\ref{complete U}) and~(\ref{rotation}), we get
\begin{align}
N=&\mathbb{I}_{3 \times 3}-\frac{1}{2}(M_D^*(M^*_R)^{-1}M^{-1}_RM^{T}_D)_{3\times 3}V_1 ,\\
S=& (M^*_D(M_R^*)^{-1})_{3\times m}V_2.
\end{align}
The nonunitary effects are related to the light sector, in other words with the N block matrix. We can parametrize the nonunitary deviation as: 
\begin{equation}
N=(\mathbb{I}-\eta)V_1,
\end{equation}
where $\eta$ describes the nonunitary effects in terms of the $M_D$ and $M_R$, 
\begin{equation}
\eta=\frac{1}{2}(M_D^*(M^*_R)^{-1}M^{-1}_RM^{T}_D)_{3\times 3}.
\end{equation}
Originally, this type-I seesaw model was motivated by Grand Unification Theories. Therefore, in that case, the NHL masses were expected at high energy scales, of the order of the GUT scale. Currently, well-motivated low-energy realizations of the seesaw mechanism are being considered since the expected mass scales might be close to the energy range of near future accelerator experiments. This is the case of the Linear seesaw mechanism that we consider in this work.

\subsection*{Linear seesaw mechanism}
In several low-scale realizations of the type-I seesaw mechanism, such as the linear seesaw, instead of adding only three singlets under $SU(2)$, the number of additional NHL is extended to be six. The mass matrix in the linear seesaw has the form~\cite{Malinsky:2005bi}:

\begin{equation}
\label{linear seesaw mechanism}
 M_{\nu}=\begin{pmatrix}
0 & M_D & M_L\\ 
M^T_D & 0 & M\\
M^T_L & M^T  & 0
\end{pmatrix} .
\end{equation}
All the components in this matrix are $3\times 3$
  block matrices.  Our purpose is to analyze the lepton flavor violation phenomenology of
this well motivated scheme. 
The most important requirement in our analysis is to respect the hierarchy 
$M_L<<M_D<<M$. 

To diagonalize 
Eq.~(\ref{linear seesaw mechanism}), it is usual  to define the
following matrices:
\begin{align}
\label{Masses linear}
M_{DL}=& (M_D,M_L)_{3\times 6} \quad
M^T_{DL}= \begin{pmatrix}
    M^T_D \\
    M^T_L
\end{pmatrix}_{6 \times 3},\\ \notag \\ \label{masses linear 2}
M_{R_{6 \times 6}}=& \begin{pmatrix} 
0 & M \\
M^T & 0
\end{pmatrix} \quad
(M_{R_{6 \times 6}})^{-1}= \begin{pmatrix} 
0 & (M^T)^{-1} \\
M^{-1} & 0
\end{pmatrix}
\end{align} 
and diagonalize by blocks to get the light and heavy masses 
\begin{equation}
m_{\nu}= M_D(M_LM^{-1})^T +(M_LM^{-1})M^T_D, \quad M_N= M_{R_{6 \times 6}}.
\end{equation}
We have that the rectangular matrix, equivalent to Eq.~(\ref{kappa}), has the form
\begin{equation}
\label{K neutrino}
    K^{LS}=(N^{LS}_{3\times3},S^{LS}_{3 \times 6}),
\end{equation}
where (LS) is for Linear Seesaw. In the limit $M_L \longrightarrow 0$, the submatrices take the form
\begin{equation}
\label{K components}
    N^{LS}=(\mathbb{I}_{3\times 3}-\eta^{LS}_{3\times 3})\cdot V_{1_{3\times 3}}, \quad S^{LS}=(0_{3\times 3},(M^*_D(M^{*T})^{-1})_{3\times 3})\cdot V_{2_{6\times 6}} 
\end{equation}
and the nonunitary effects are characterized by: 
\begin{equation}
\label{nonunitary matrix}
\eta^{LS}=\frac{1}{2}(M_D^*(M^*)^{-1}M^{-1}M^{T}_D)_{3\times 3}.
\end{equation}

\subsection*{Matching the non-unitary effects in the different parametrizations}
The mixing matrix that we see above comes from the type-I seesaw mechanism. However, it is not the only realization of the mixing matrix, an example is the symmetric parametrization. Therefore, we have another way to describe the non-unitary effects in the light sector and we want to match the $\eta$ parameters with the non-unitary parameters on this parametrization. 
The relevance of the symmetric parametrization lies in the constraints on the non-unitary parameters obtained from oscillation experiments \cite{Chatterjee:2021xyu,Celestino-Ramirez:2023zox,CentellesChulia:2024sff}.  In this parametrization, each mixing angle has a Majorana phase. For instance, 
\begin{equation}
\omega_{13}=\begin{pmatrix}
\cos \theta_{13} & 0 & \sin \theta_{13} e^{-i\phi_{13}}\\ 
0 & 1 & 0\\
-\sin \theta_{13} e^{i\phi_{13}}  & 0 & \cos \theta_{13}
\end{pmatrix}.
\end{equation} 
The standard $3\times3$ leptonic mixing matrix is written in this case as
$U^{SM} = \omega_{23}\omega_{13}\omega_{12}$. We can extend the matrix as
many new neutrinos as we want, and the light sector is described as
follows\footnote{An equivalent triangular parametrization for fixed number of extra neutral heavy leptons was discussed in Refs.~\cite{Xing:2007zj,Xing:2011ur}}~\cite{Escrihuela:2015wra}:
 \begin{equation}
\label{N}
    N=N^\mathrm{NP}U^{SM}=\begin{pmatrix}
\alpha_{11} & 0 & 0 \\ 
\alpha_{21} & \alpha_{22} &0 \\
\alpha_{31} &  \alpha_{32} & \alpha_{33}
\end{pmatrix} U^{SM},   
\end{equation}
 where $N^{NP}$ parametrize the deviation from the unitary in the light sector. Since the $3\times 3$ matrix, $N$, is no longer unitary, it will have up to 15 independent parameters: three mixing angles and three phases contained in the $U^{SM}$, three non-diagonal non-unitary $\alpha_{ij}$ parameters (with $i\neq j$) with its three phases, and three real $\alpha_{ii}$ diagonal parameters. This parametrization is important in constraining non-unitary effects in oscillation experiments.
 
The non-unitary effect in the symmetrical parametrization is a complete description of the light sector and it can be compared with the parametrization in terms of the $\eta$ matrix. We can also obtain an expression in terms of the extra mixing angles when we consider that they are small.
For this purpose, we approximate the mixing matrix to quadratic terms in the mixing angle that are related to the active neutrinos and the NHLs. For example, to describe explicitely the mixing matrix in the $3+3$ case we have
\begin{eqnarray}
  U^{6x6} &=& D(\beta) \omega_{56}\omega_{46}\omega_{36}\omega_{26}\omega_{16}\omega_{45}\omega_{35}\omega_{25}\omega_{15}\omega_{34}\omega_{24}\omega_{14}\omega_{23}\omega_{13}\omega_{12}, \nonumber \\ 
 &=& D(\beta) U^{NHL} U^{3x3}
\end{eqnarray}
This unitary matrix has $6^2$ parameters, although many of them will not be physical.  Here, $D(\beta)$ is a diagonal matrix\footnote{It will be shown below that this matrix could be considered as containing part of the non-physical phases and, therefore, could be taken as the unit matrix.} $diag(e^{i\beta_1},e^{i\beta_2},e^{i\beta_3},...e^{i\beta_n})$ and the product of the last three matrices
$U^{3x3} = \omega_{23}\omega_{13}\omega_{12}$ describes the mixing of the light
neutrino sector with the three standard active neutrinos. On the other hand, the first products describe the mixing of the light and heavy sector with the additional NHLs 
\begin{equation}
  \label{U_heavy}
  U^{NHL}=\omega_{56}\omega_{46}\omega_{36}\omega_{26}\omega_{16}\omega_{45}\omega_{35}\omega_{25}\omega_{15}\omega_{34}\omega_{24}\omega_{14}.
\end{equation}
This particular ordering for the multiplication of the matrices gives
rise to the triangular parametrization described by the submatrix
$N^{NP}$.  We can go a step forward and rearrange this matrix to
separate it into two different parts without loosing the triangular
property for $N^{NP}$. We first notice that the matrix $\omega_{45}$
conmutes with the product $\omega_{36}\omega_{26}\omega_{16}$ since the mixings acts on different rows and columns. Due to this property, we can rearrange the matrix in Eq.~(\ref{U_heavy}), without changing parametrization, as 
\begin{equation}
  U^{NHL} =\omega_{56}\omega_{46}\omega_{45}\omega_{36}\omega_{26}\omega_{16}\omega_{35}\omega_{25}\omega_{15}\omega_{34}\omega_{24}\omega_{14}
\end{equation}
Notice that the matrices that mix the heavy sector are now grouped: 
\begin{equation}
  U^H=\omega_{56}\omega_{46}\omega_{45}.
\end{equation}
On the other hand,  those that mix the additional NHLs with the light neutrino states appears to the right of $U^{NHL}$:  
\begin{equation}
  U^{LH}=\omega_{36}\omega_{26}\omega_{16}\omega_{35}\omega_{25}\omega_{15}\omega_{34}\omega_{24}\omega_{14}.\end{equation}
In this way, we can separate the complete $n\times n$ mixing matrix
into the product of three different matrices that contain the
information of the purely heavy sector, $U^{H}$, the purely light
sector, $U^{3\times3}$, and the mixing of both sectors, $U^{LH}$,
respectively. Notice that this separation can be applied
irrespectively of the number of extra NHLs. We restrict
here to the $6\times 6$ case but the same method applies in general.

The full mixing matrix in this $6\times 6$ example, or in the $n\times n$ case, can then be written as
\begin{equation}
\label{unitarymatrix}
   U^{6\times 6}= D(\beta)U^{H}U^{LH}U^{3\times3} = D(\beta )\begin{pmatrix}
I_{3\times 3} & 0_{3\times 3}  \\ 
0_{3\times 3} & H_{3\times 3}  
\end{pmatrix} U^{LH}\begin{pmatrix}
U^{SM} & 0_{3\times 3}  \\ 
0_{3\times 3} & I_{3\times 3}  
\end{pmatrix}.
\end{equation}
Where $U^{SM}$ and $H_{3\times 3}$ are unitary matrices. Concentrating ourselves on the first three rows of this matrix, that are the ones with the physical information, we can notice that the first three phases in $D(\beta)$ can be absorbed by the three charged lepton fields. Besides, we can see the deviation of the unitary in the light sector comes from the $U^{LH}$ matrix. After this rearrangement, it is evident that the submatrix $N^{NP}$ will not have any contribution from the "heavy-heavy'' mixing angles and will depend only on angles (and phases) of the form $\theta_{1n},\theta_{2n},\theta_{3n}$, with $n>3$, while the $U^{SM}$ submatrix will depend on the angles (and phases) $\theta_{12}$, $\theta_{13}$, and $\theta_{23}$ ($\phi_{12}$, $\phi_{13}$, and $\phi_{23}$). That is, defining $N_S$ as the number of additional NHL, we will have $3+3N_S = 3(n-2)$ mixings and $3+3N_S = 3(n-2)$ phases, as expected~\cite{Giunti:2007ry,Rodejohann:2011vc,Valle:2015pba}.

Using the approximation where we consider the new mixing angles up to order 2, the light sector of $U^{LH}$ is~\footnote{This $6\times 6$ case parametrization had already been studied in~\cite{Xing:2007zj,Xing:2011ur}}:
\begin{equation}
\label{approximation}
    \begin{pmatrix}
\alpha_{11} & 0 & 0 \\ 
\alpha_{21} & \alpha_{22} &0 \\
\alpha_{31} &  \alpha_{32} & \alpha_{33}
\end{pmatrix}
\approx I_{3\times 3}-\begin{pmatrix}
\frac{1}{2}(\theta^2_{14}+\theta^2_{15}+\theta^2_{16}) & 0 & 0 \\ 
\hat{\theta}_{16}\hat{\theta}^{*}_{26}+\hat{\theta}_{15}\hat{\theta}^{*}_{25}+\hat{\theta}_{14}\hat{\theta}^{*}_{24} & \frac{1}{2}(\theta^2_{24}+\theta^2_{25}+\theta^2_{26}) &0 \\
\hat{\theta}_{16}\hat{\theta}^{*}_{36}+\hat{\theta}_{15}\hat{\theta}^{*}_{35}+\hat{\theta}_{14}\hat{\theta}^{*}_{34} & \hat{\theta}_{26}\hat{\theta}^{*}_{36}+\hat{\theta}_{25}\hat{\theta}^{*}_{35}+\hat{\theta}_{24}\hat{\theta}^{*}_{34}  & \theta^2_{14}+\theta^2_{15}+\theta^2_{16}
\end{pmatrix}.
\end{equation}
Where $\hat{\theta}_{ij}=\theta_{ij}e^{i\phi_{ij}}$. 
We can generalize this approximated matrix  for the case of an arbitrary number of new neutral heavy leptons: 
\begin{equation}
\begin{pmatrix}
\sum_{j=4}^{n}\frac{1}{2}(\theta^2_{1j}) & 0 & 0 \\ 
\sum_{j=4}^{n}\hat{\theta}_{1j}\hat{\theta}^{*}_{2j} & \sum_{j=4}^{n}\frac{1}{2}(\theta^2_{2j})&0 \\
\sum_{j=4}^{n}\hat{\theta}_{1j}\hat{\theta}^{*}_{3j} &\sum_{j=4}^{n}\hat{\theta}_{2j}\hat{\theta}^{*}_{3j}  & \sum_{j=4}^{n}\frac{1}{2}(\theta^2_{3j})
\end{pmatrix}=
\begin{pmatrix}
\alpha^{\prime}_{11} & 0 & 0 \\ 
\alpha^{\prime}_{21} & \alpha^{\prime}_{22} &0 \\
\alpha^{\prime}_{31} &  \alpha^{\prime}_{32} & \alpha^{\prime}_{33}
\end{pmatrix}=\alpha^{\prime}_{3\times 3},
\end{equation}
where n is the total number of neutrinos. The particular case of $9\times 9$ has  been studied in Ref.~\cite{Han:2021qum}.

To find the equivalence between the two different parametrizations, we compute the expression for $NN^{\dagger}$ in both cases: 
\begin{align}
NN^{\dagger}&=(\mathbb{I}-\eta)(\mathbb{I}-\eta)^{\dagger}=\mathbb{I}-2\eta +\mathcal{O}(2)\\
NN^{\dagger}&=(\mathbb{I}-\alpha^{\prime})(\mathbb{I}-\alpha^{\prime})^{\dagger}=\mathbb{I}-(\alpha^{\prime}+\alpha^{\prime^{\dagger}}) +\mathcal{O}(2)\\
\label{matching}
\eta&=\frac{1}{2}(\alpha^{\prime}+\alpha^{\prime^{\dagger}})=\frac{1}{2}\begin{pmatrix}
2\alpha^{\prime}_{11} & \alpha^{\prime^*}_{21} & \alpha^{\prime^*}_{31} \\ 
\alpha^{\prime}_{21} & 2\alpha^{\prime}_{22} &\alpha^{\prime^*}_{32} \\
\alpha^{\prime}_{31} &  \alpha^{\prime}_{32} & 2\alpha^{\prime}_{33}
\end{pmatrix} ,
\end{align}
in accordance with \cite{Blennow:2016jkn}. Now we have a relation between both parametrization and constraints for $\eta$ elements are constraints to $\alpha$ parameters.

\section{cLFV and numerical analysis}
Heavy neutral leptons within the type I seesaw framework contribute to lepton flavor violation processes due to their mixing with the active SU(2) doublet neutrinos. We will focus on processes of the type  $\ell_i \longrightarrow \ell_{j}\gamma$ and $\mu^{-} \to e^{-}e^{-}e^{+}$, where $\ell_{i}$ could be a $\tau$ or a $\mu$, and $\ell_{j}$ stands for e or $\mu$. In general, these branching ratios are given by\footnote{See also~\cite{Ilakovac:1994kj}}~\cite{He:2002pva}

\begin{equation}
\label{branching}
Br(\ell_i \longrightarrow \ell_{j}\gamma)= \frac{\alpha^3_Ws^2_W}{256\pi^2}\frac{m^5_{\ell_1}}{M^4_{W}}\frac{1}{\Gamma_{\ell_{i}}}|G^W_{ij}|^2,   
\end{equation}
where
\begin{align}
G^W_{ij}=& \sum_{k=1}^{9} K^{*}_{ik}K_{jk}G^W_{\gamma}\left(\frac{m^2_i}{M_W^2}\right),\notag \\ 
G^W_{\gamma}(x)=&\frac{1}{12(1-x)^4}(10-43x+78x^2-49x^3+18x^3lnx+4x^4),
\end{align}
$M_W$ is the boson W mass, $m_i$ is the physical neutrino mass, and
$\alpha_W \equiv \alpha / s^2_W$ with $\alpha = e^2 /4\pi$ the fine
structure constant and $s_W$ the Weinberg angle. We will analyze the nonunitary effects
  using the current constraints and the future sensitivity of these
  processes, summarized in Table~\ref{table 1}, as reported in~\cite{Garnica:2023ccx}.

\begin{table}[]
\def\arraystretch{2}
\begin{tabular}{|c|c|c|}
\hline
Process & Present limit & Future Sensitivity \\ \hline
$\mu \longrightarrow e \gamma$       & $4.2 \times 10^{-13}$ \cite{MEG:2013oxv}         & $6 \times 10^{-14}$ \cite{MEGII:2018kmf}                  \\ \hline
$\tau \longrightarrow e \gamma$      & $3.3 \times 10^{-8}$ \cite{BaBar:2009hkt}          & $3 \times 10^{-9}$\cite{Belle-II:2018jsg}                  \\ \hline
$\tau \longrightarrow \mu \gamma$       & $4.2 \times 10^{-8}$ \cite{BaBar:2009hkt}         & $10^{-9}$  \cite{Belle-II:2018jsg}                \\ \hline
$\mu^{-} \to e^{-}e^{-}e^{+}$ & $10^{-12}$ \cite{SINDRUM:1987nra} & $10^{-16}$ \cite{Hesketh:2022wgw,Mu3e:2020gyw} \\ \hline
\end{tabular}
\caption{\label{table 1} The current limits and expected future sensitivity for the flavor violations processes, }
\end{table}
The branching ratio for the $\mu^{-} \to e^{-}e^{-}e^{+}$ process is \cite{Ilakovac:1994kj}:  
\begin{align}
 Br(\mu\to e^{-}e^{-}e^{+})=& \frac{\alpha^{4}_{w}}{24576 \pi^{3}}\frac{m^4_{\mu}}{M^4_W}\frac{m_{\mu}}{\Gamma_{\mu}} \times \bigg(2|\frac{1}{2}F^{\mu eee}_{Box}+F^{\mu e}_{Z}-2s^2_w(F^{\mu e}_{Z}-F^{\mu e}_{\gamma})|^2+4s^4_w|F^{\mu e}_{Z}-F^{\mu e}_{\gamma}|^2 \notag \\
 & +16 s^2_wRe\left[(F^{\mu e}_{Z}+\frac{1}{2}F^{\mu eee}_{Box})\times(G^{\mu e}_{\gamma})^*\right]-48s^4_wRe\left[(F^{\mu e}_{Z}-F^{\mu eee}_{Box})\times(G^{\mu e}_{\gamma})^*\right] \notag \\
 &+32s^4_w Re|G^{\mu e}_{\gamma}|^2 \left[\ln\left(\frac{m^2_{\mu}}{m^2_{e}}\right)-\frac{11}{4}\right]\bigg),
\end{align}
where
\begin{align}
F^{\mu eee}_{Box} &= 2\sum_{ij} K^{*}_{\mu i}K^{*}_{ej} K_{ei}K_{ej}F_{Box}(\lambda_i, \lambda_j)+\sum_{ij}K^{*}_{\mu i}K^{*}_{ei}K_{ej}K_{ej}G_{Box}(\lambda_i,\lambda_j), \\
F^{\mu e}_Z &= \sum_{ij} K^{*}_{\mu e}K_{ej}\left[\delta_{ij}F_Z(\lambda_i)+C^{*}_{ij}G_z(\lambda_i,\lambda_j)+C_{ij}H_z(\lambda_i,\lambda_j)\right], \\ 
G^{\mu e}_{\gamma} &=\sum_i K^{*}_{\mu i}K_{ei}G_{\gamma}(\lambda_{i}), \\
G^{\mu e}_{\gamma} &=\sum_i K^{*}_{\mu i}K_{ei}F_{\gamma}(\lambda_{i}),
\end{align}
where $\lambda_i=\frac{m^2_{\nu_i}}{M^2_W}$ and 
\begin{align}
F_{\gamma}(x=\lambda_i) &= \frac{7x^3-x^2-12x}{12(1-x)^3}-\frac{x^4-10x^3+12x^2}{6(1-x)^4}\ln x,  \\ 
G_{\gamma}(x) &= -\frac{2x^3 +5x^2-x}{4(1-x)^3}- \frac{3x^2}{2(1-x)^4}\ln x, \\
F_Z(x) &=-\frac{5x}{2(1-x)}-\frac{5x^2}{2(1-x)^2}\ln x, \\
G_Z(x,y) &= - \frac{1}{2(x-y)}\left[\frac{x^2(1-y)}{1-x}\ln x-\frac{y^2(1-x)}{1-y}\right], \\
H_Z(x,y)&= \frac{\sqrt{xy}}{4(x-y)}\left[\frac{x^2 -4x}{1-x}\ln x -\frac{y^2-4y}{1-y}\ln y  \right], \\ 
F_{Box}(x,y)&=\frac{1}{x-y}\bigg[(1+\frac{xy}{4})\big(\frac{1}{1-x}+ \frac{x^2\ln x}{(1-x)^2}-\frac{1}{1-y}-\frac{y^2 \ln y}{(1-y)^2}\big) \notag \\
&-2xy\big(\frac{1}{1-x}\big)+\frac{x \ln x}{(1-x)^2}-\frac{1}{1-y}-\frac{y\ln y}{(1-y)^2}\bigg], \\ 
G_{Box}(x,y) &= -\frac{\sqrt{xy}}{x-y}\bigg[(4+xy)\big(\frac{1}{1-x}+\frac{x\ln x}{(1-x)^2}-\frac{1}{1-y}-\frac{y\ln y}{(1-y)^2}\big) \notag \\
&-2\big(\frac{1}{1-x}+\frac{x^2\ln x}{(1-x)^2}-\frac{1}{1-y}-\frac{y^2\ln y}{(1-y)^2}\big)\bigg], \\
C_{ij}&=\sum_{k=1}^{3}K^{*}_{l_k i}K_{l_k j}
\end{align}
Finally for the case that $x=y$ we use the next functions
\begin{align}
 G_Z(x,x) &= -\frac{x}{2}-\frac{x\ln x}{1-x}, \\
 H_Z(x,x) &= \frac{3}{4}-\frac{x}{4}-\frac{3}{4(1-x)}, \\
 F_{Box}(x,x) &= -\frac{x^4-16x^3+19x^2-4}{4(1-x)^3}-\frac{3x^3+4x^2-4x}{2(1-x)^3} \ln x, \\
 G_{Box}(x,x) &= \frac{2x^4-4x^3+8x^2-6x}{(1-x)^3}-\frac{x^4+x^3+4x}{(1-x)^3}\ln x
\end{align}
To reduce the number of degrees of freedom of our analysis, we will work in a scheme where $M_L$ and $M$ are diagonal and real. We will assume that the flavor violation is only due to the Dirac Yukawa coupling~\cite{DAmbrosio:2002vsn}. To write the Dirac mass matrix, we use the Casas-Ibarra parametrization applied to the linear seesaw~\cite{Forero:2011pc,Casas:2001sr}.

\begin{equation}
M_D=V^*_1 \sqrt{m^{diag}_{\nu}}A^T \sqrt{m^{diag}_{\nu}}V^{*T}_1(M^T_L)^{-1}M^T,
\end{equation}
where A takes the form:
\begin{equation}
A=\begin{pmatrix}
\frac{1}{2} & -a & -b\\ 
a & \frac{1}{2} & -c\\
b & c  & \frac{1}{2}
\end{pmatrix}.
\end{equation}
Here a,b, and c are real. In this scheme, $V_1$ is the unitary $3\times 3$ leptonic mixing matrix, and $m^{diag}_{\nu}$ is a diagonal matrix of the light physical neutrino masses. We define the matrices $M_L$ and M as 
\begin{align}
M=& v_{M}diag(1+\epsilon_{M_{11}},1+\epsilon_{M_{22}},1+\epsilon_{M_{33}}), \\
M_L=&v_{L}diag(1+\epsilon_{L_{11}},1+\epsilon_{L_{22}},1+\epsilon_{L_{33}}),
\end{align}
where $v_{M}$ and $v_{L}$ are the vacuum expectation values (vev). We fix the to $v_{M}=1$~TeV, and for $v_{L}$ we consider values in the range of $[10^{-1}-10^2]$ eV.
\begin{table}[]
\def\arraystretch{2}
\begin{tabular}{|c|c|c|}
\hline
Parameters & Normal ordering at 3$\sigma$ & Inverse ordering at 3$\sigma$ \\ \hline
       $\Delta m^2_{21}$($eV^2$)    & $(6.94-8.14)\times 10^{-5}$  & $(6.94-8.14)\times 10^{-5}$                     \\ \hline
$\Delta m^2_{31}$($eV^2$)           & $(2.47-2.63)\times 10^{-3}$   &  $(2.37-2.53)\times 10^{-3}$                  \\ \hline
$\theta_{12}/^{\circ}$           &   31.4-37.4     &    31.4-37.4           \\ \hline
   $\theta_{23}/^{\circ}$        & 41.2-51.33       &            41.16-51.25   \\ \hline
$\theta_{13}/^{\circ}$           & 8.13-8.92        &         8.17-8.96     \\ \hline
    $\delta/^{\circ}$       &   128-359           &    200-353     \\ \hline
\end{tabular}
\caption{\label{oscillation parameters} Allowed parameter range, at $3 \sigma$~\cite{deSalas:2020pgw} level, for the mixing angles, CP-violating phase, and neutrino mass differences. We use these range for the scan of the $V_1$ matrix.}
\end{table}

  We perform a random scan of the free parameters involved in the lepton flavor violation processes. 
  For the parameters of the $3\times 3$ leptonic mixing matrix, we use the current values at the $3 \sigma$~\cite{deSalas:2020pgw} confidence level, shown in Table~\ref{oscillation parameters}. Notice that this range for the mixing angles gives variations above the percentage level, that will be bigger than the variations in the non-unitary parameters; that is, the violation of non-unitarity that we are considering lies inside the current experimental error. For the case of the lightest neutrino mass, we use the current cosmological constraints~\cite{deSalas:2020pgw}. Therefore, in this case, $m_1\leq 0.04$ eV. The six parameters $\epsilon_{M_{ii}}$ and $\epsilon_{L_{jj}}$, are varied in the range [-0.5,0.5]. Finally, the $a$,$b$, and $c$ parameters are varied in the range $(0-10^{-2}]$. These parameters can not be arbitrarily big due to the perturbativity condition $M_D<175$GeV. 
To constrain the non-unitary effects, we need to compute the branching ratio, Eq.~(\ref{branching}), that depends on the mixing matrix $K$. Therefore, we use the parametrization previously discussed for the submatrices $M_D$, $M_L$, and $M$ to compute the physical neutrino masses and the matrix diagonalizing the heavy sector.

\begin{figure}
\begin{subfigure}{0.4\textwidth}
    \includegraphics[width=\textwidth]{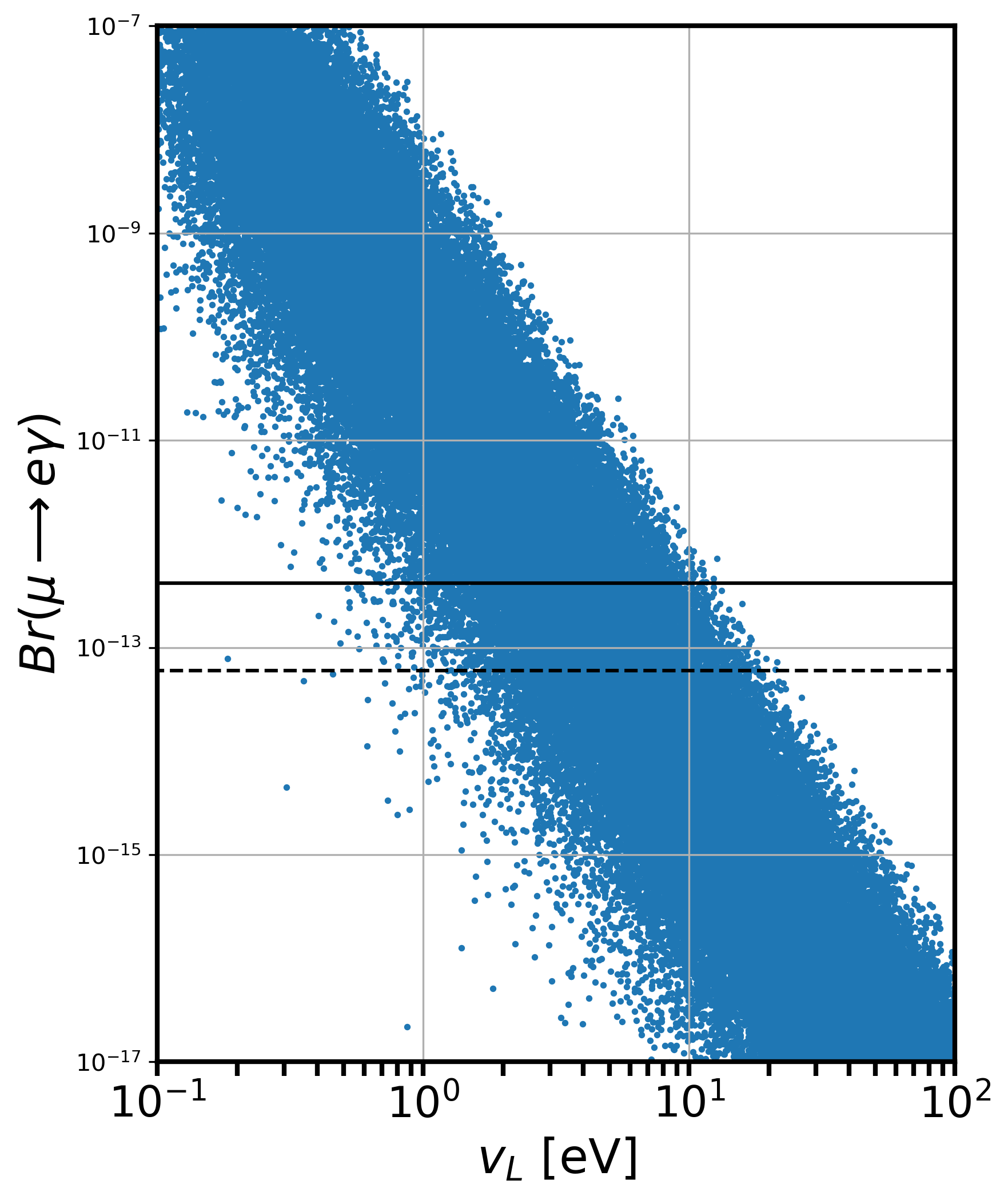}
    \label{fig:first}
    \end{subfigure}
\caption{Branching ratio of the $\mu\to e\gamma$ decay versus the flavor violation scale in eV for the normal ordering case. The solid line accounts for the current constraint, while the dashed one represents the future expected sensitivity for this decay. }
\label{fig:figure1}
\end{figure}

  \section{Resulst and Conclusions}
\begin{figure}
\begin{subfigure}{0.3\textwidth}
    \includegraphics[width=\textwidth]{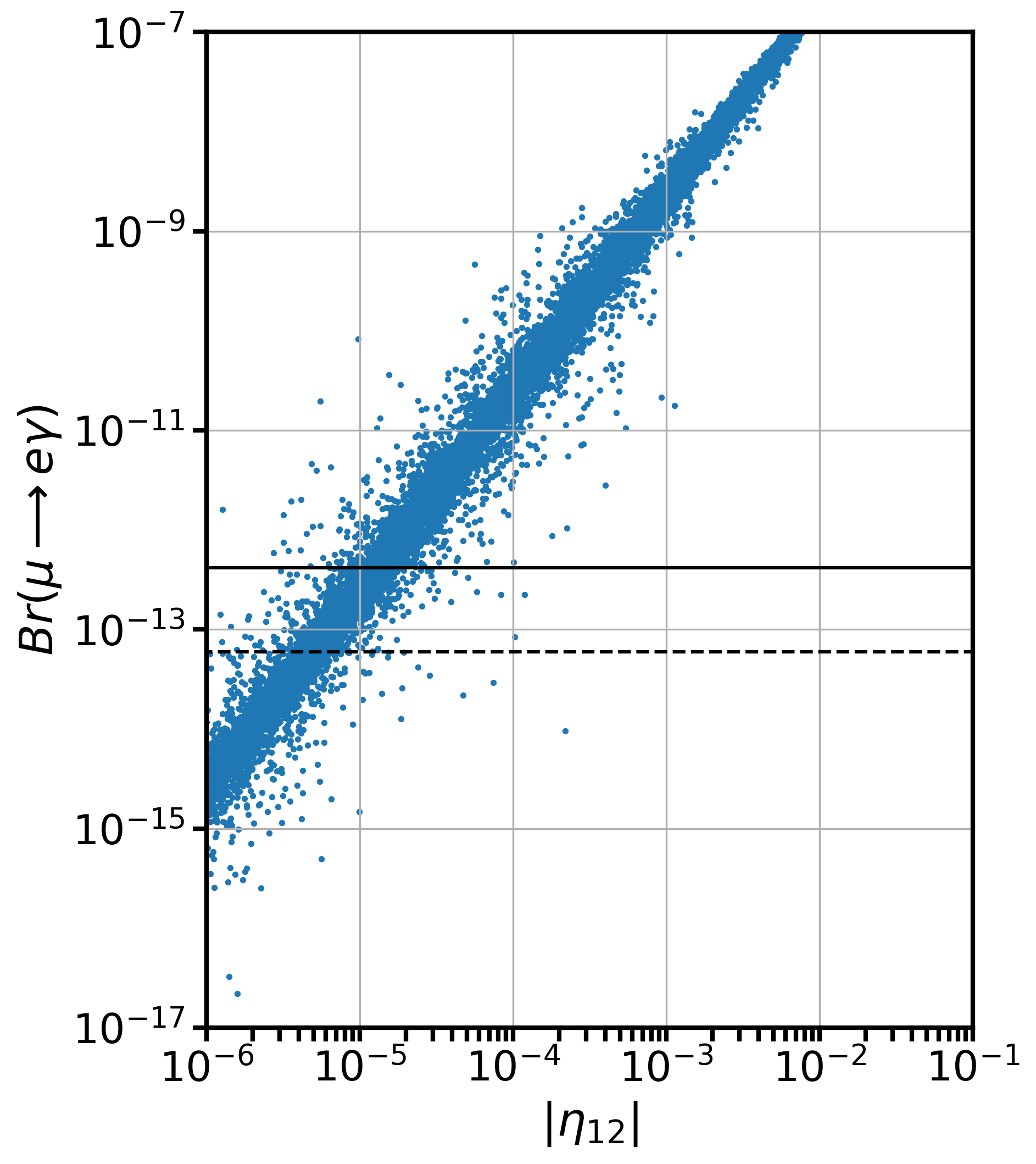}
    \end{subfigure}
    \begin{subfigure}{0.3\textwidth}
    \includegraphics[width=\textwidth]{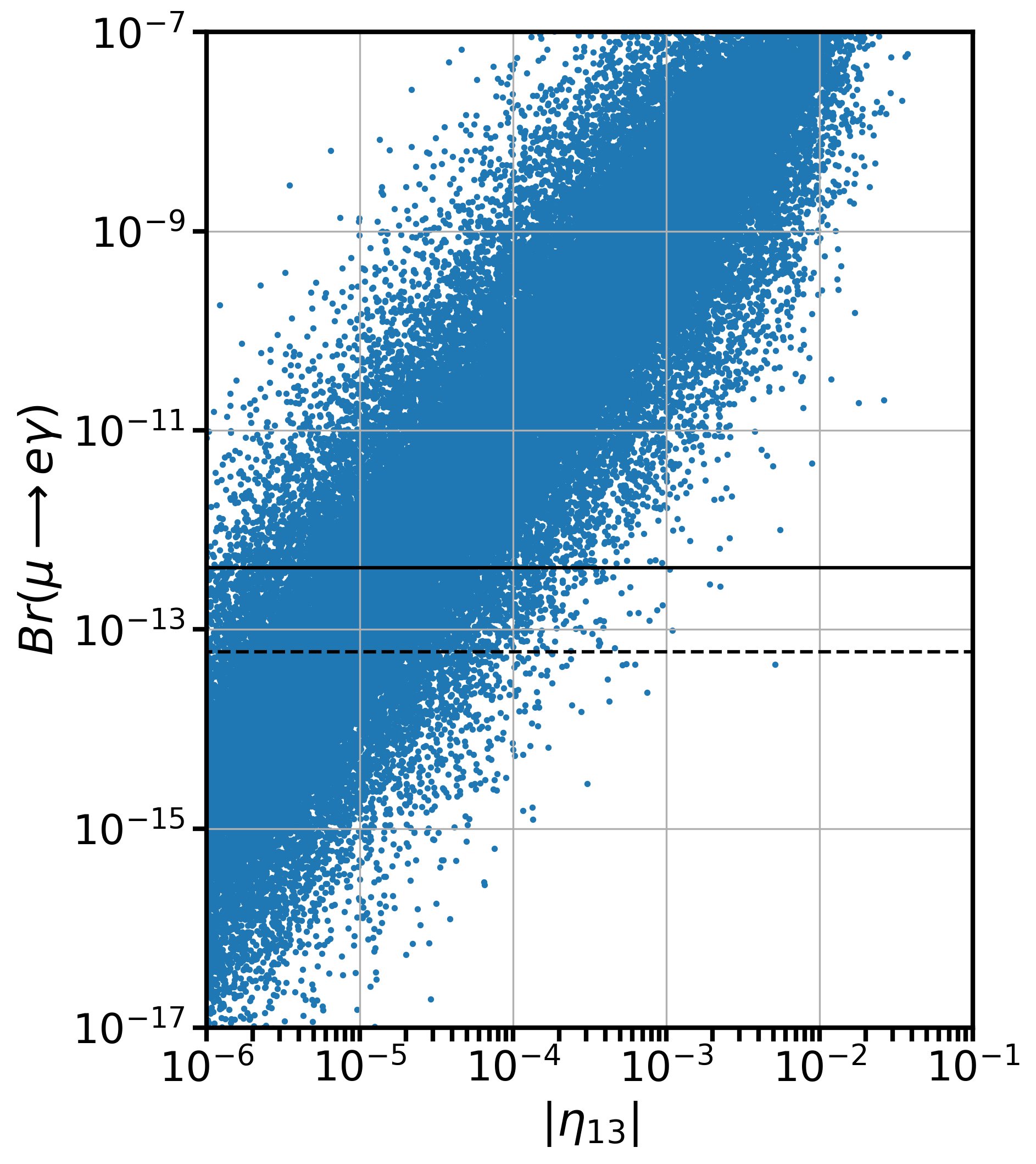}
    \end{subfigure}
\begin{subfigure}{0.3\textwidth}
    \includegraphics[width=\textwidth]{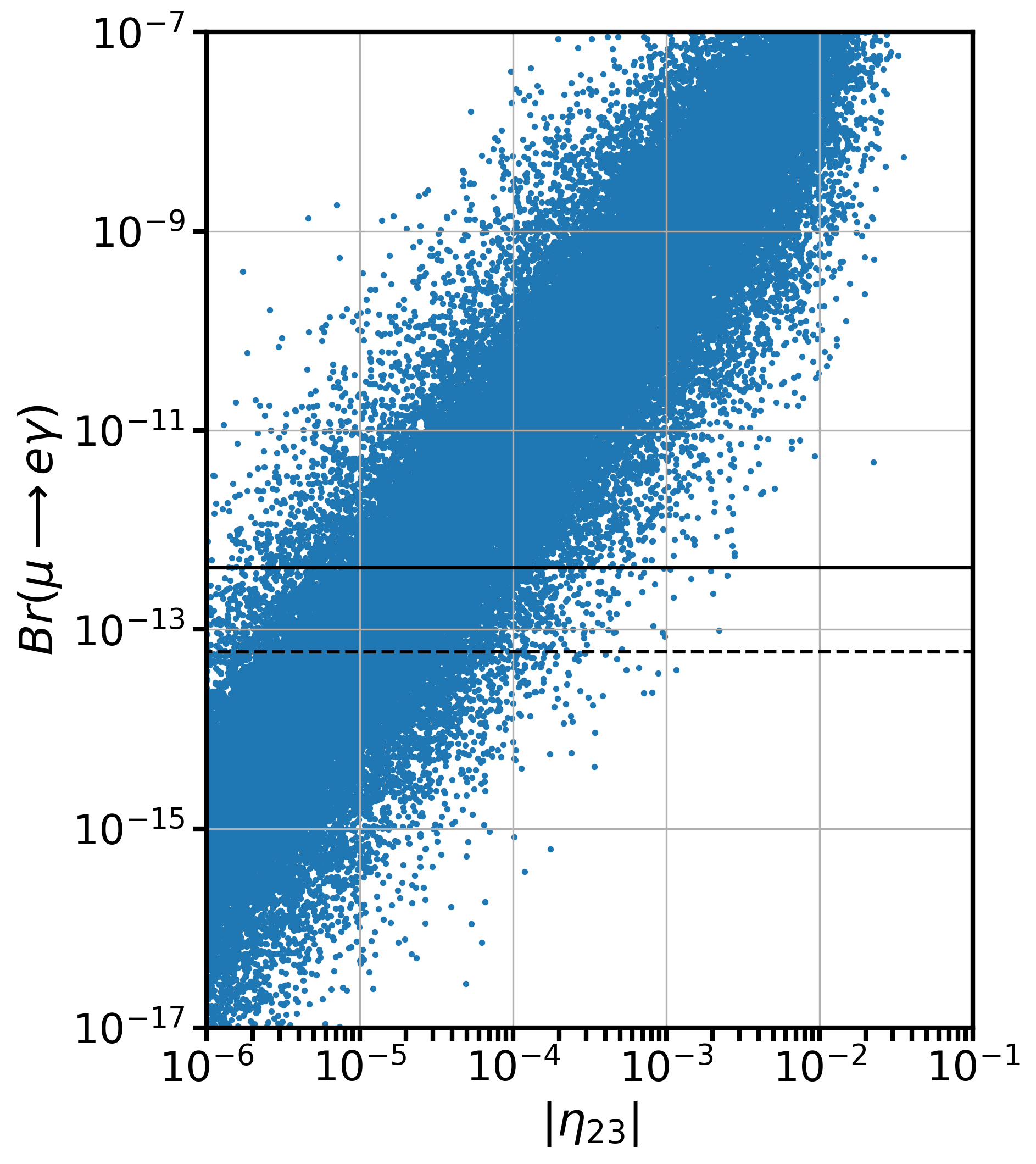}
    \end{subfigure}   
    \begin{subfigure}{0.3\textwidth}
    \includegraphics[width=\textwidth]{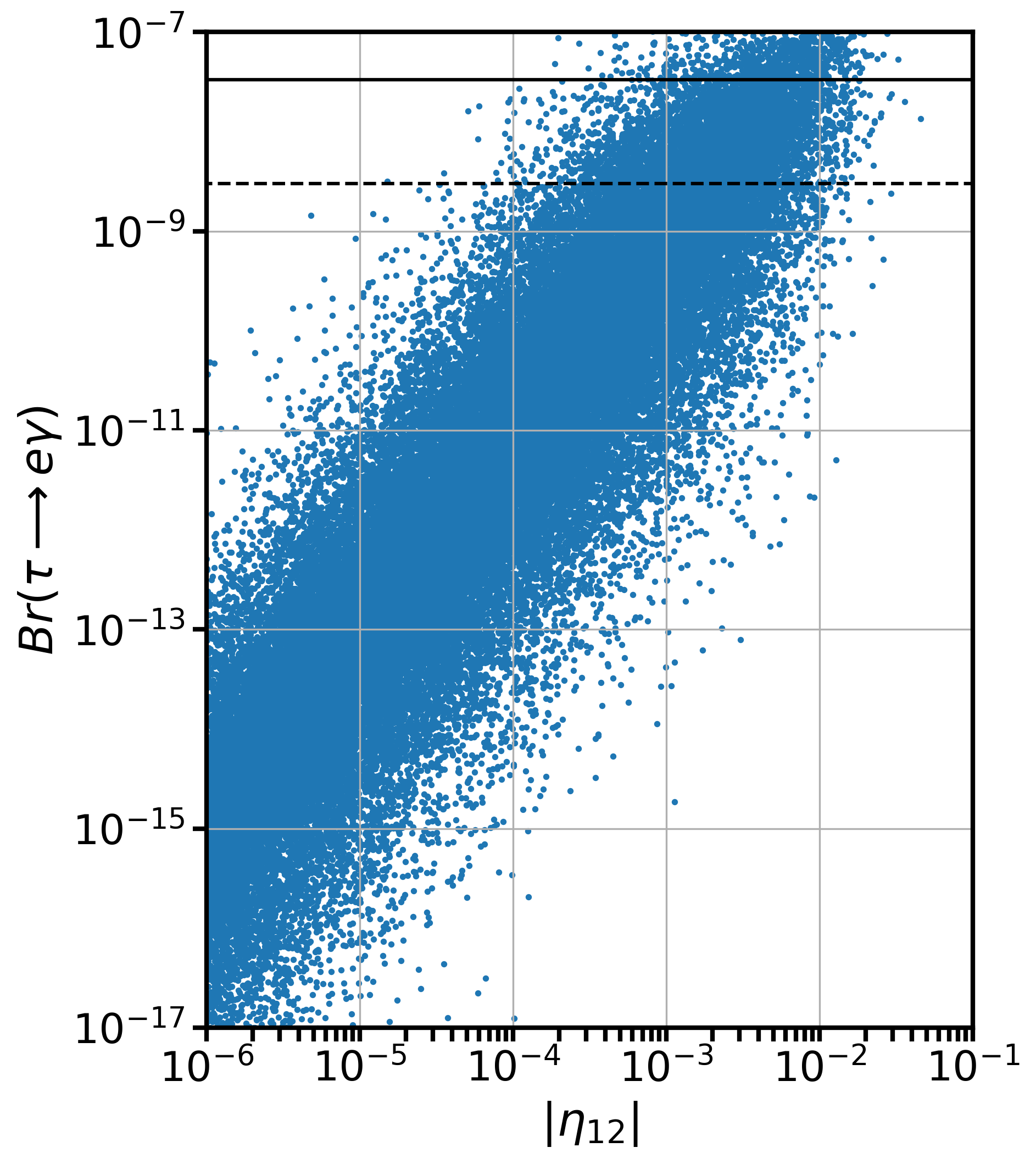}
    \end{subfigure}
    \begin{subfigure}{0.3\textwidth}
    \includegraphics[width=\textwidth]{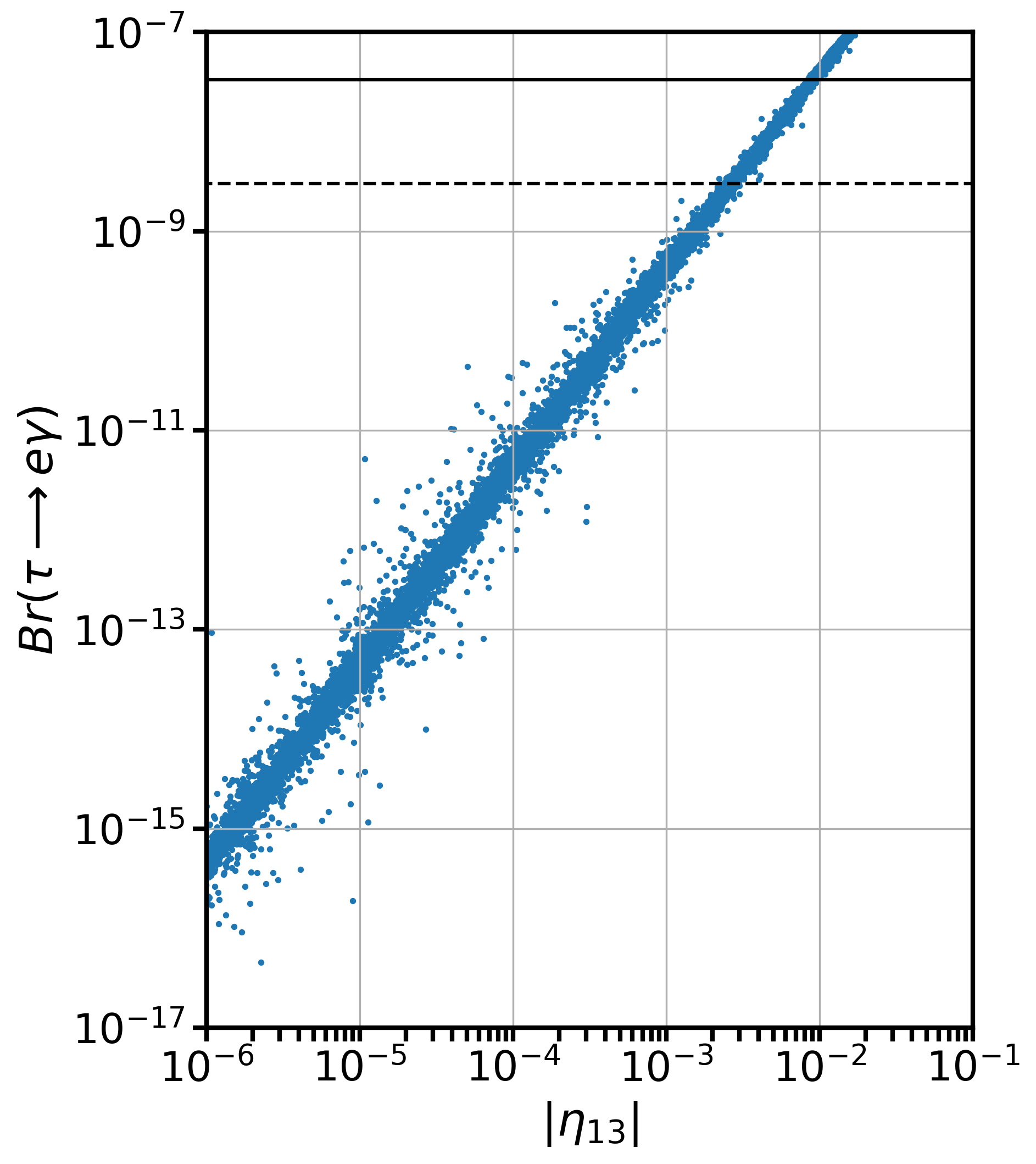}
    \end{subfigure}
\begin{subfigure}{0.3\textwidth}
    \includegraphics[width=\textwidth]{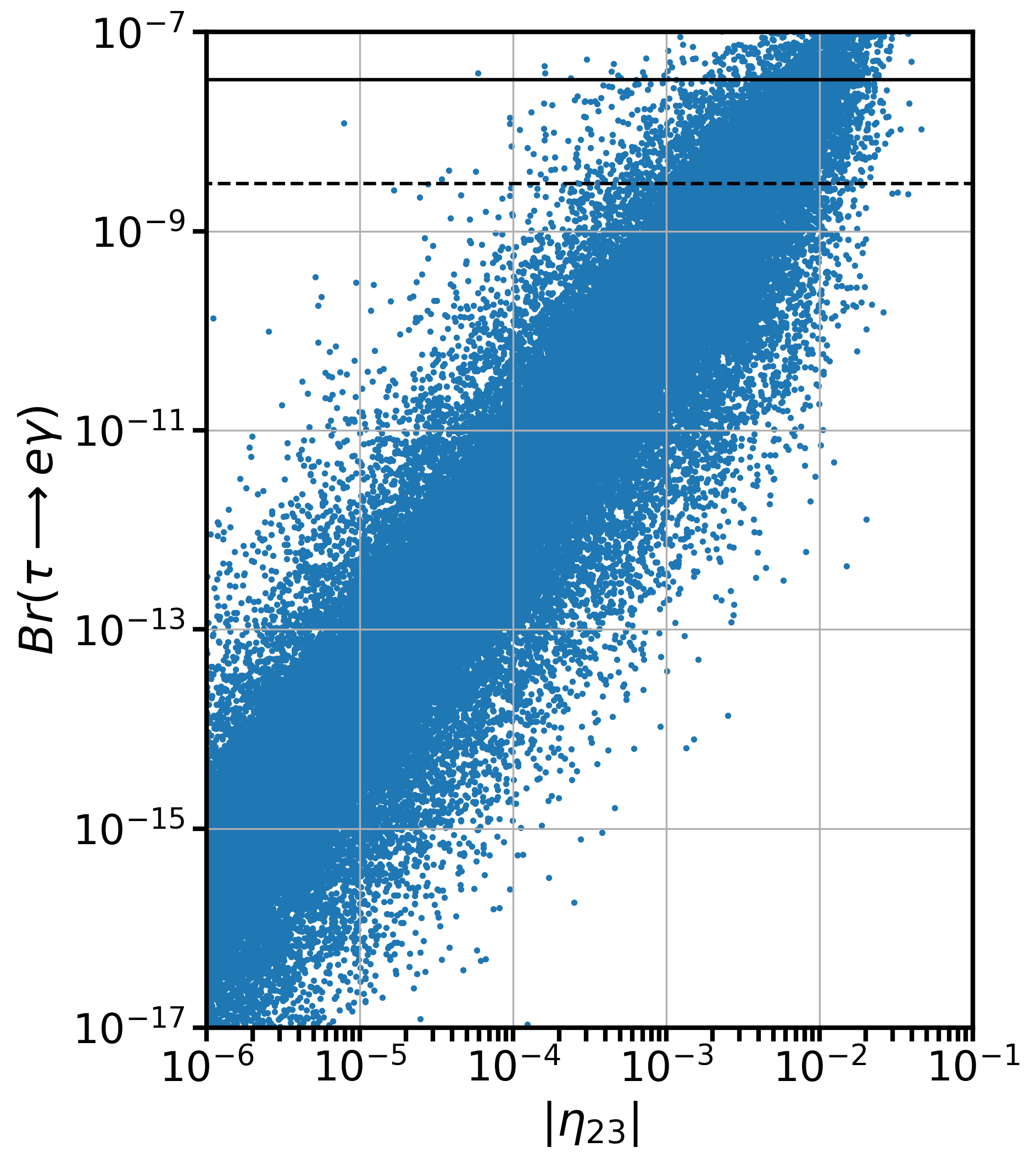}
    \end{subfigure} 
    \begin{subfigure}{0.3\textwidth}
    \includegraphics[width=\textwidth]{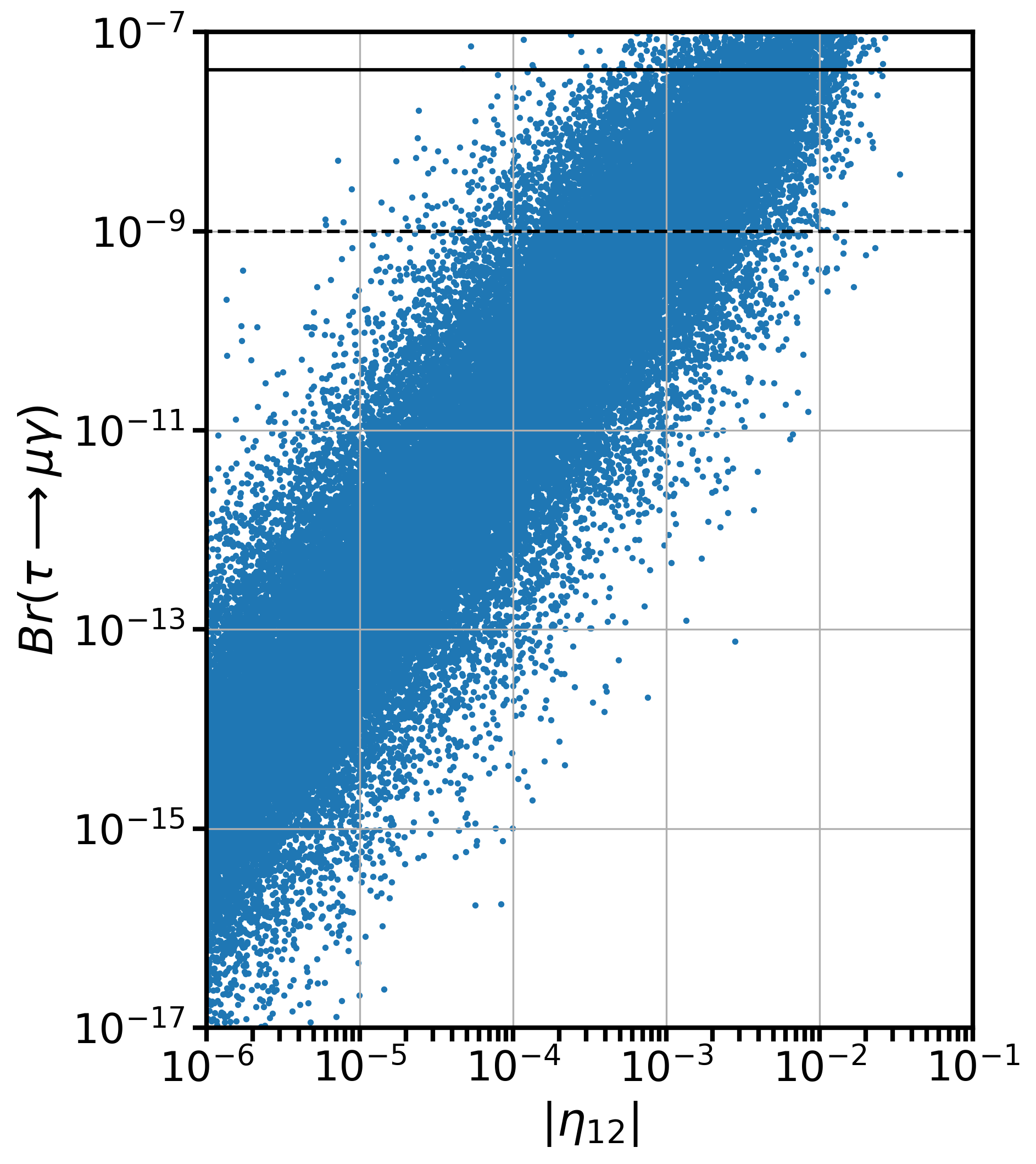}
    \end{subfigure}
    \begin{subfigure}{0.3\textwidth}
    \includegraphics[width=\textwidth]{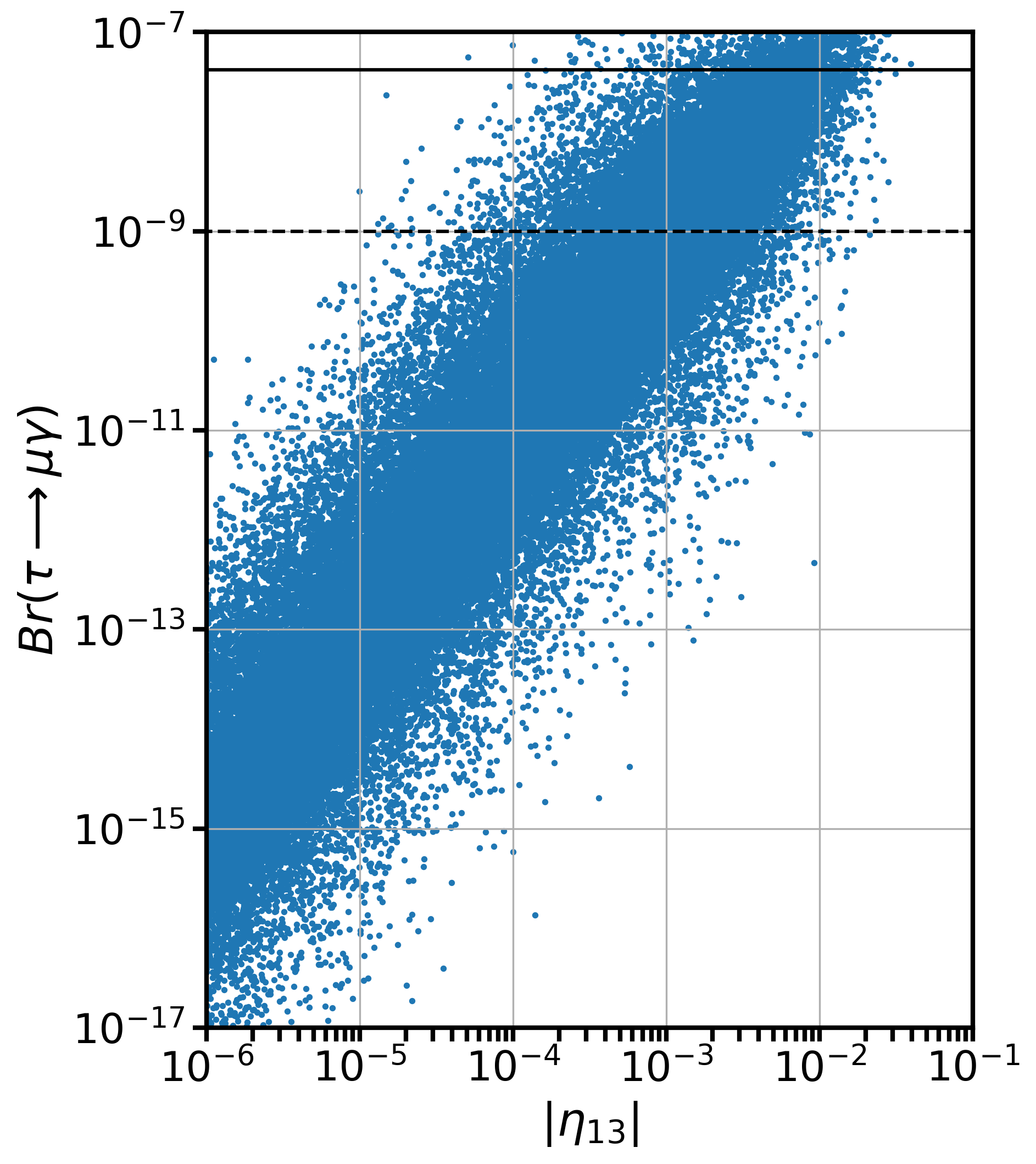}
    \end{subfigure}
\begin{subfigure}{0.3\textwidth}
    \includegraphics[width=\textwidth]{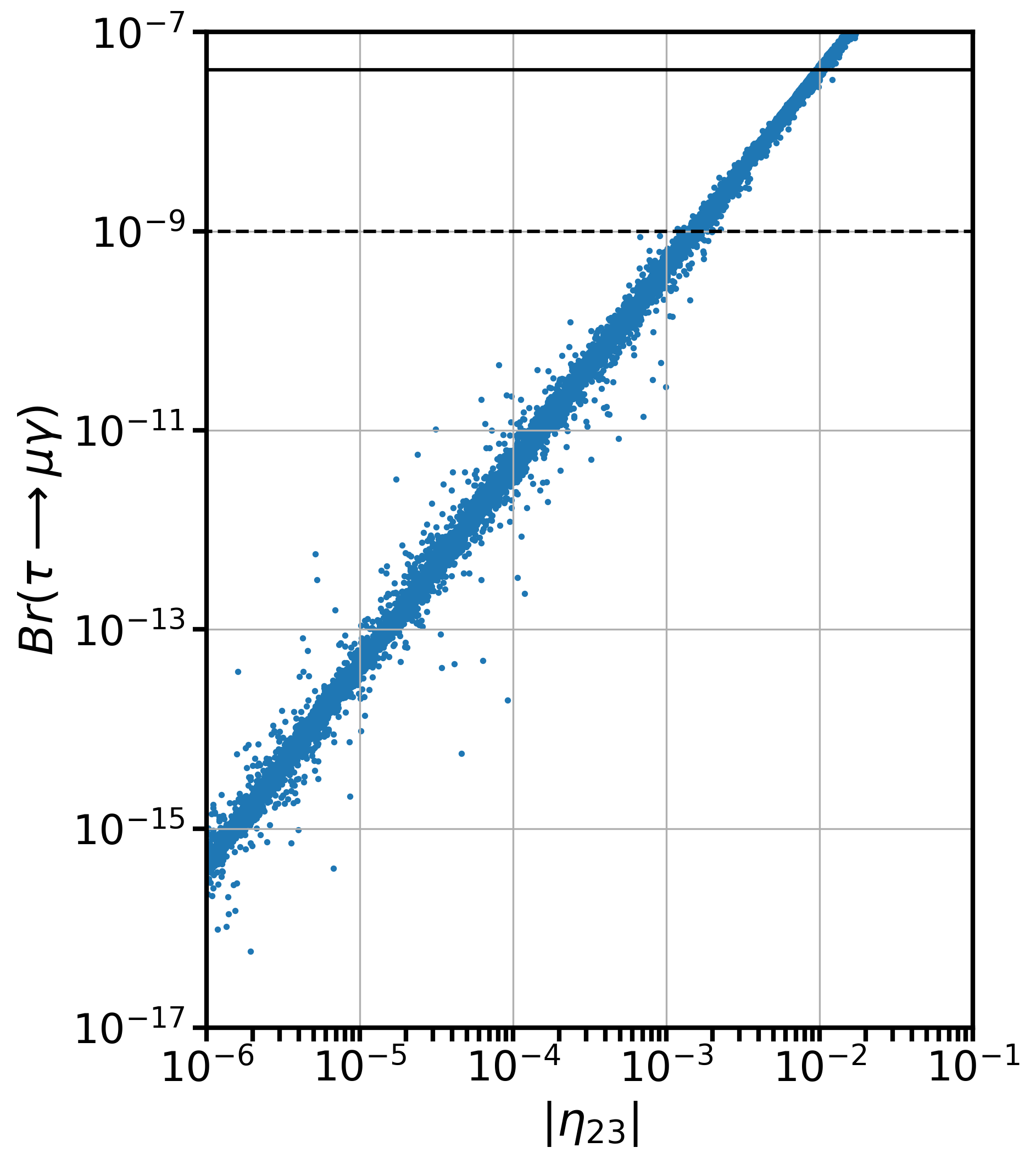}
    \end{subfigure}
\caption{\label{muon_to_electron} Scan of the branching ratio for the processes $\ell_{i} \longrightarrow \ell_{j}\gamma$ versus the nonunitary parameters in the NO case. The solid line accounts for the current constraints, while the dashed ones represent the future expected sensitivity for these decays. }
\end{figure}

\begin{figure}
\begin{subfigure}{0.3\textwidth}
    \includegraphics[width=\textwidth]{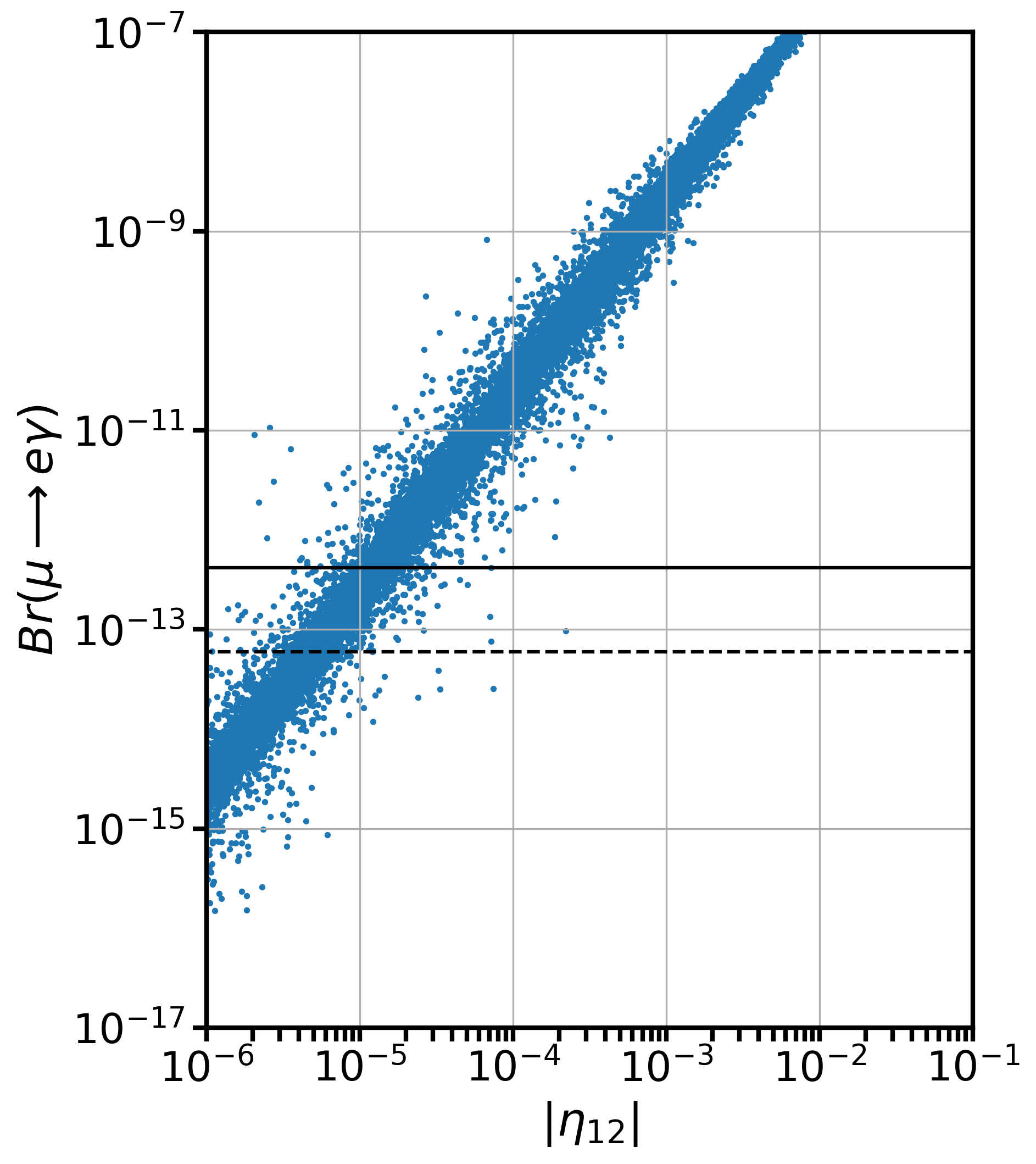}
    \end{subfigure}
    \begin{subfigure}{0.3\textwidth}
    \includegraphics[width=\textwidth]{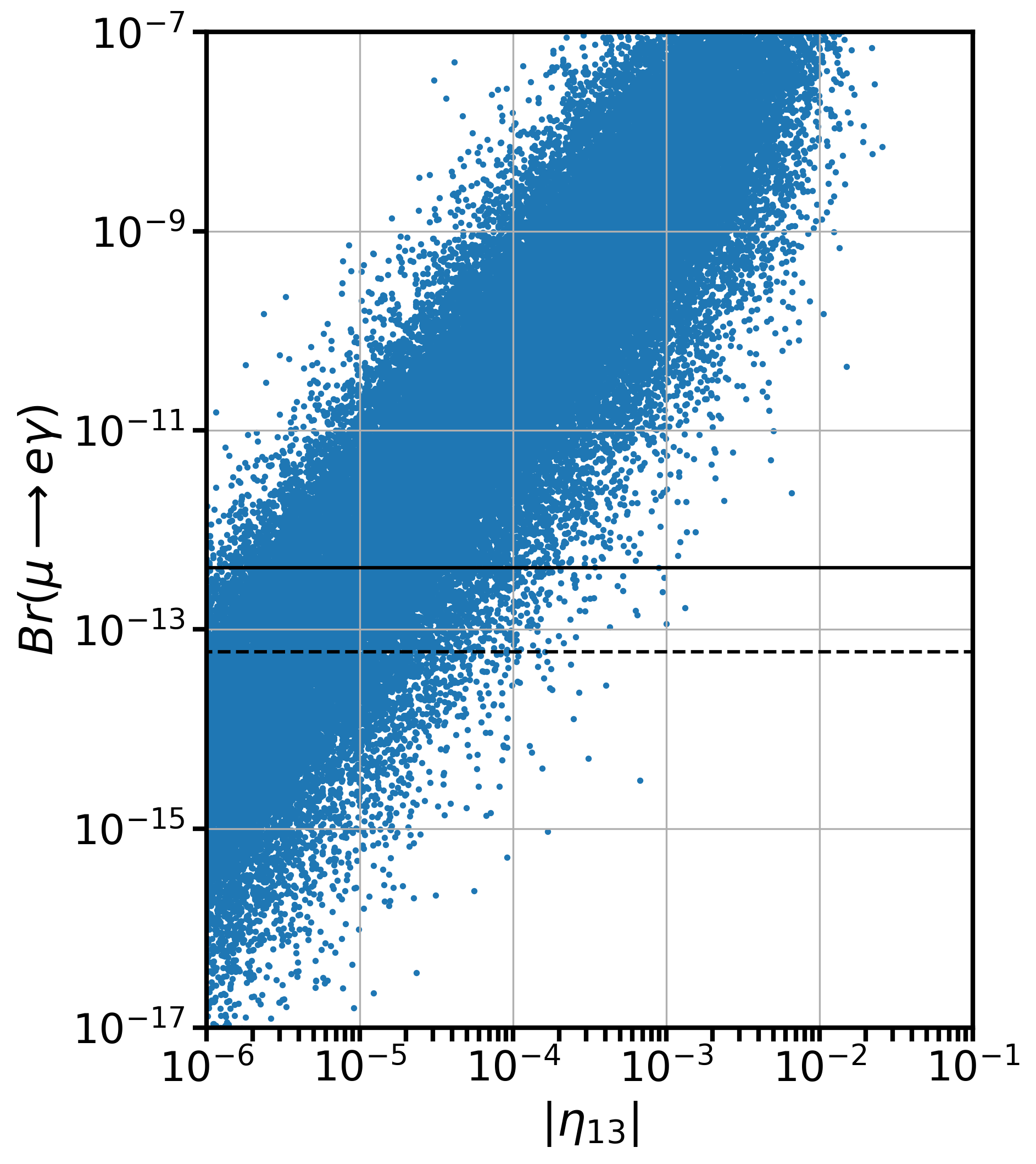}
    \end{subfigure}
\begin{subfigure}{0.3\textwidth}
    \includegraphics[width=\textwidth]{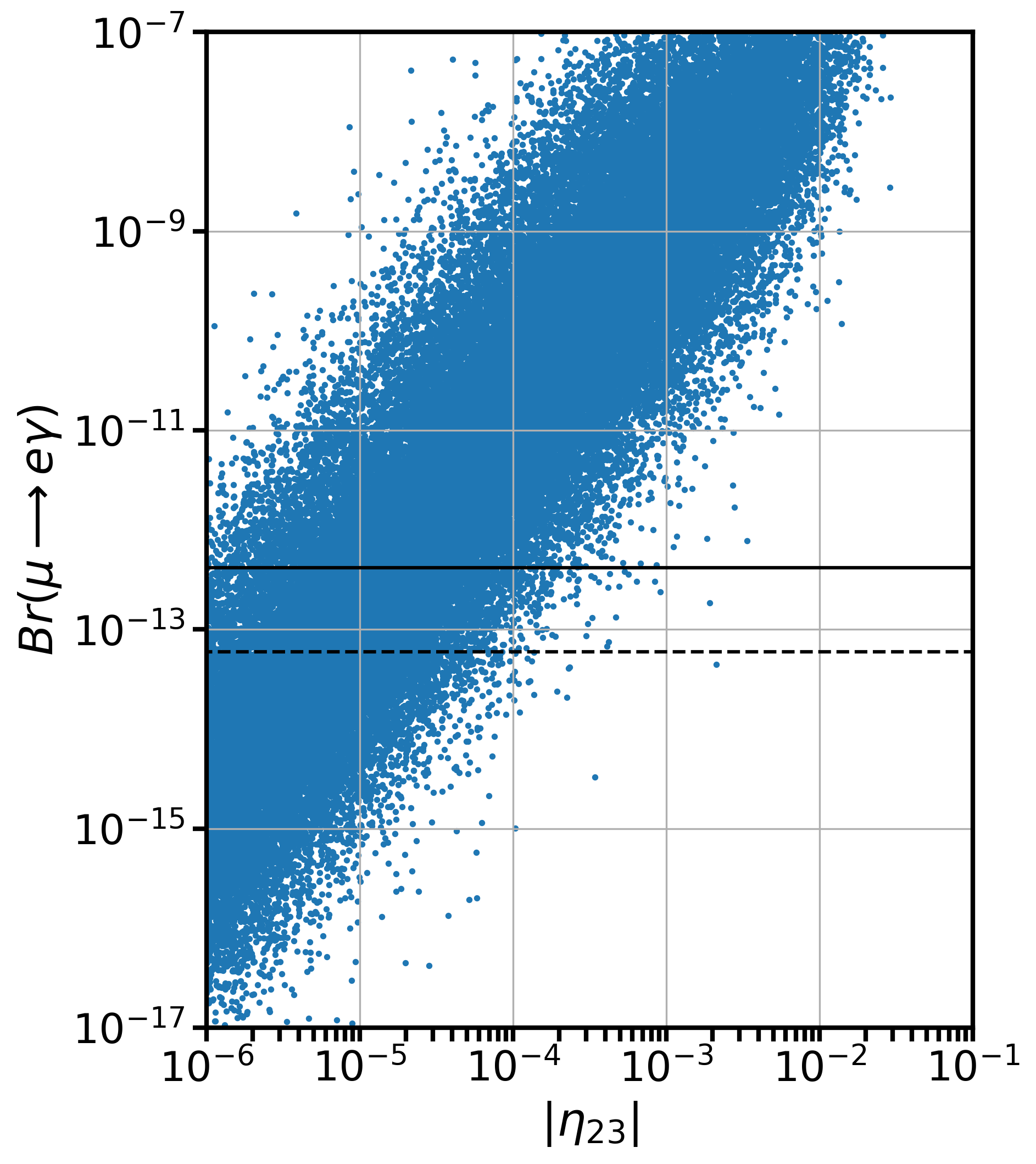}
    \end{subfigure}
    \begin{subfigure}{0.3\textwidth}
    \includegraphics[width=\textwidth]{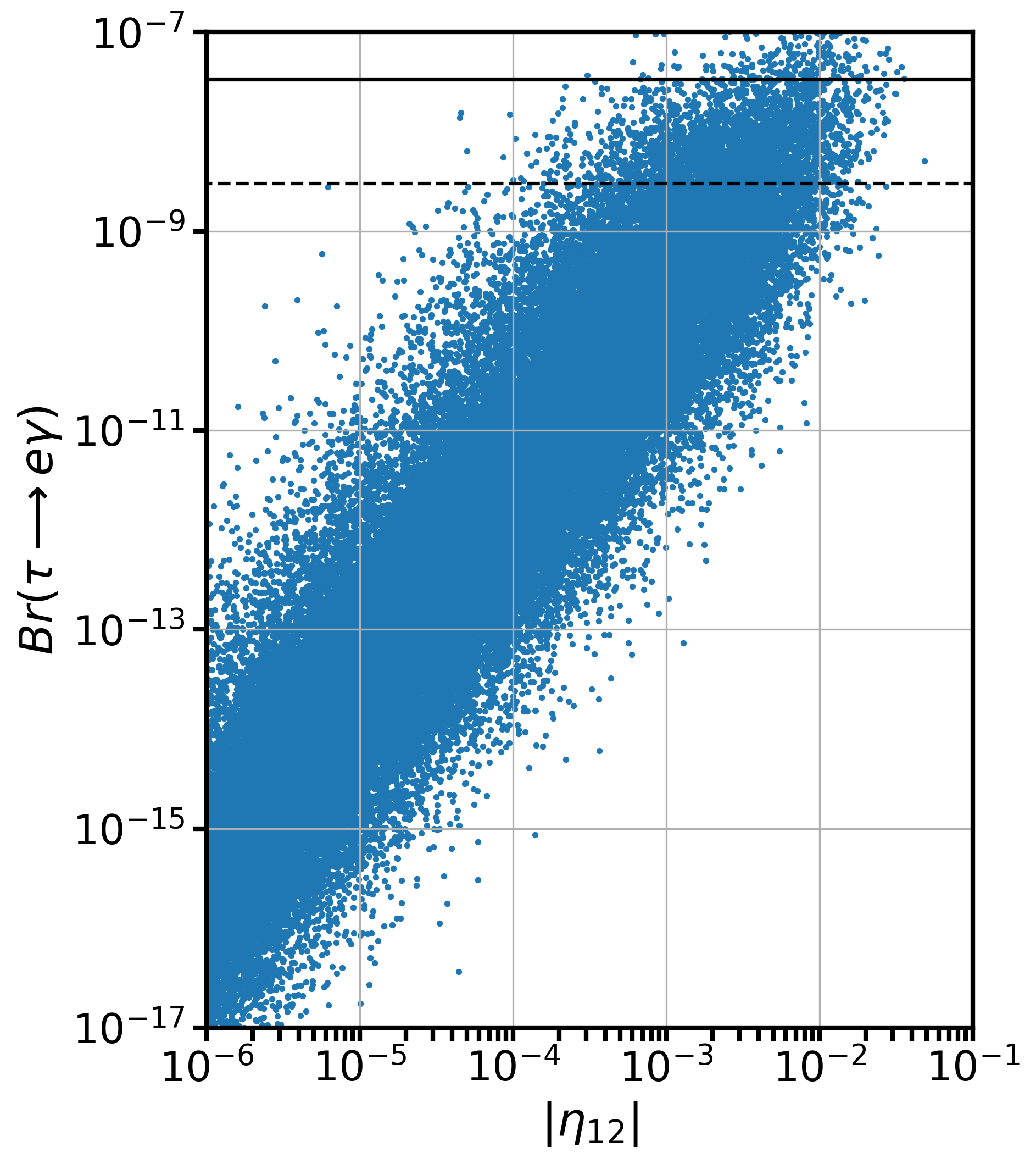}
    \end{subfigure}
    \begin{subfigure}{0.3\textwidth}
    \includegraphics[width=\textwidth]{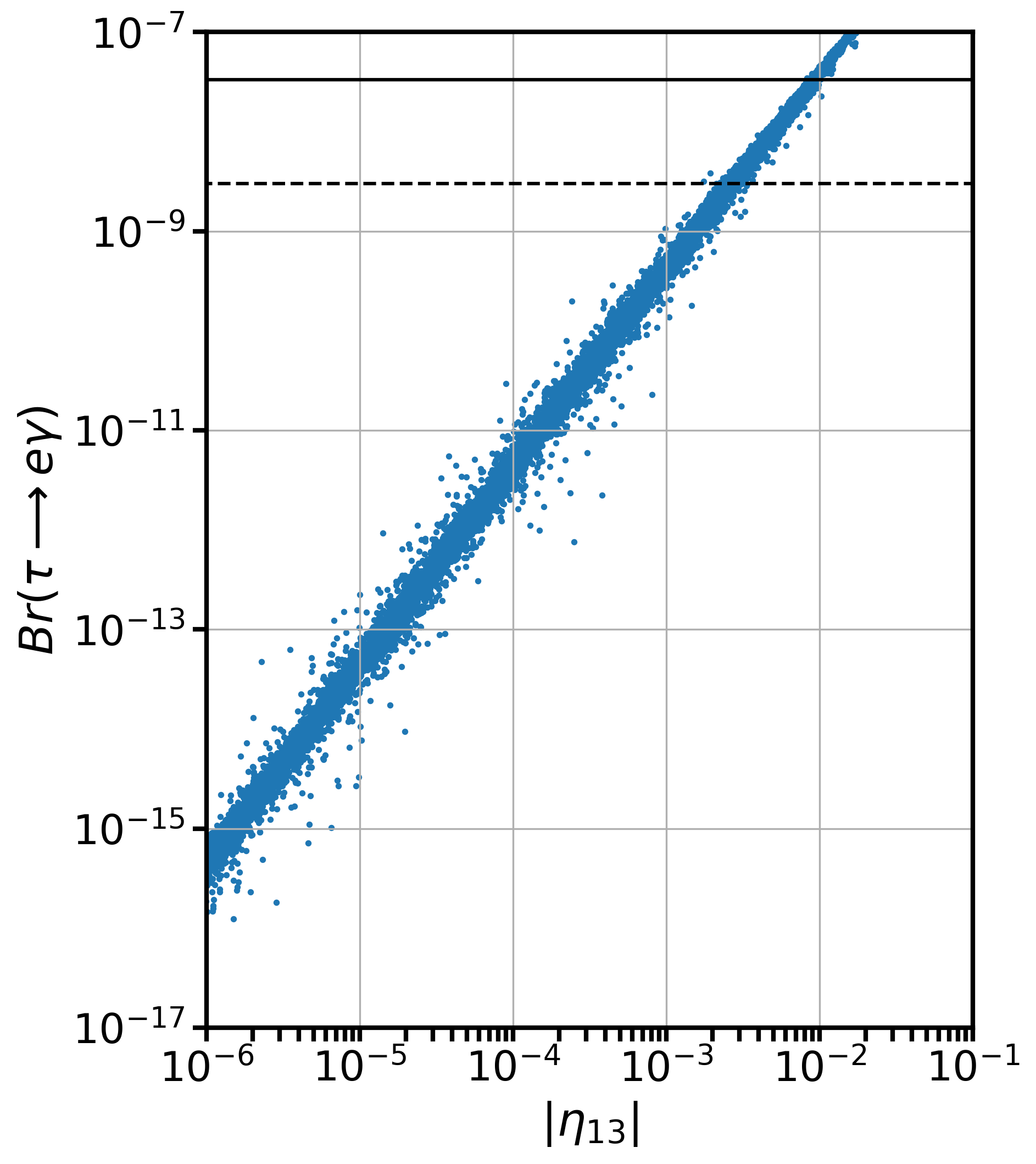}
    \end{subfigure}
\begin{subfigure}{0.3\textwidth}
    \includegraphics[width=\textwidth]{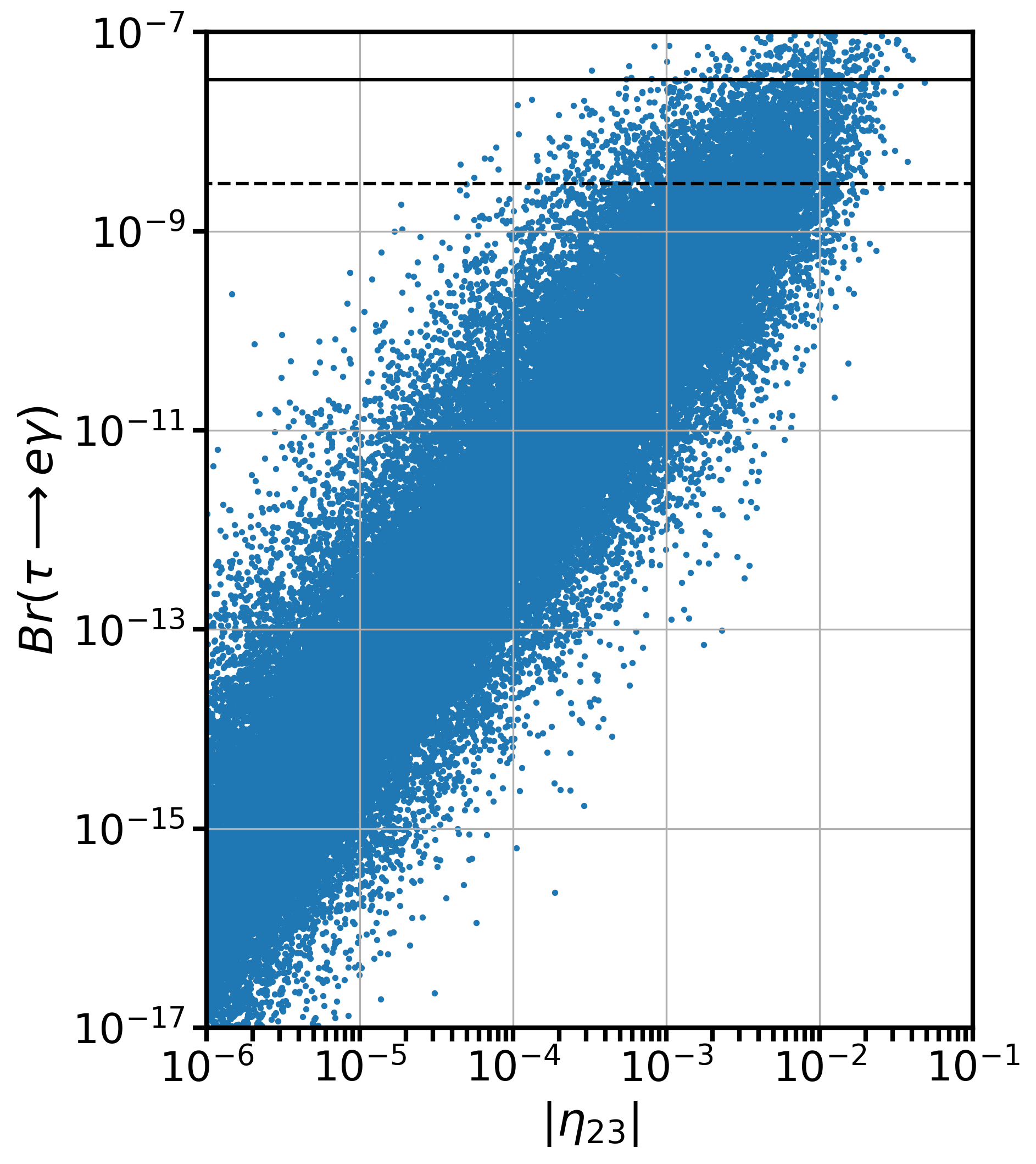}
    \end{subfigure}
\begin{subfigure}{0.3\textwidth}
    \includegraphics[width=\textwidth]{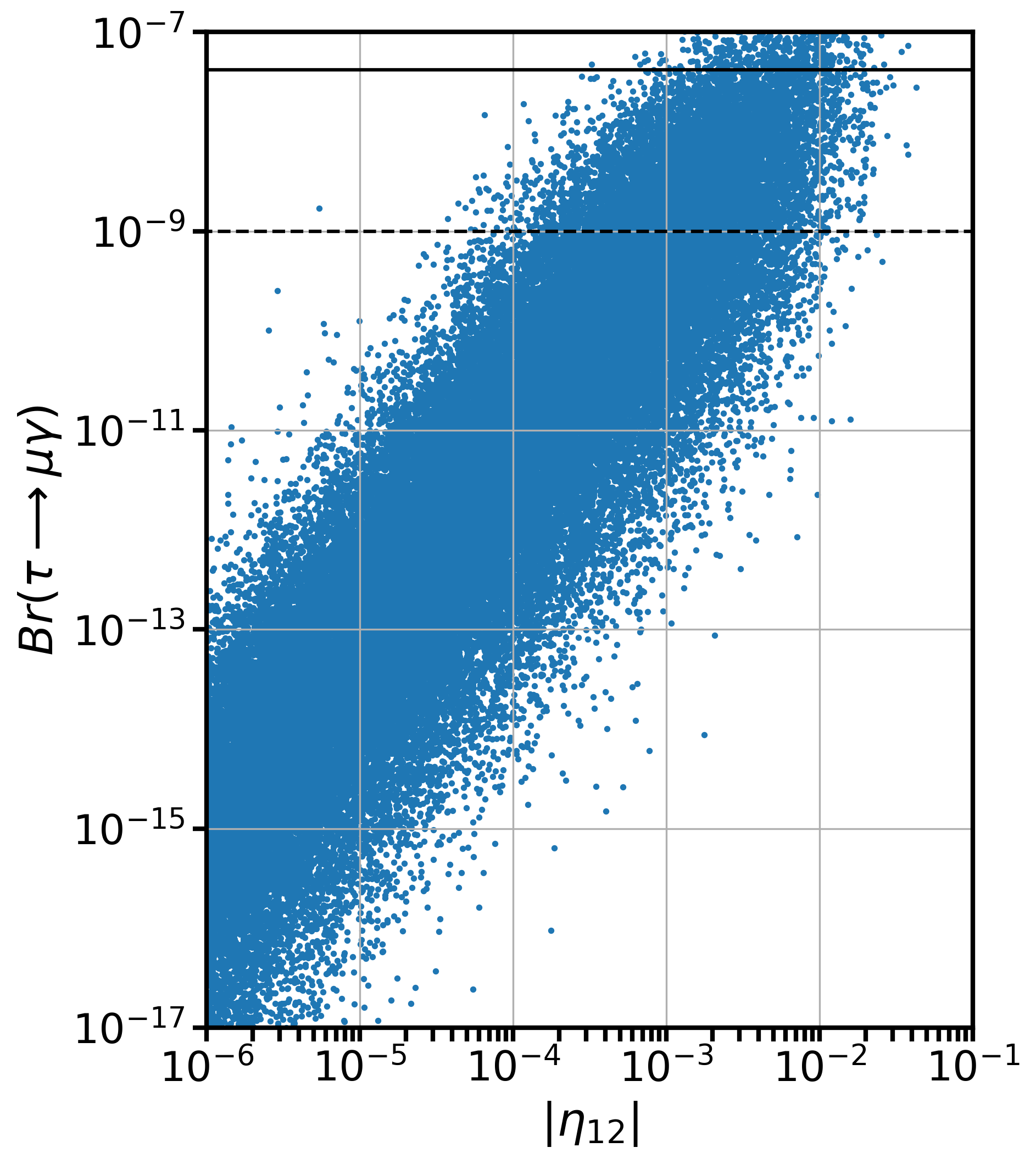}
    \end{subfigure}
    \begin{subfigure}{0.3\textwidth}
    \includegraphics[width=\textwidth]{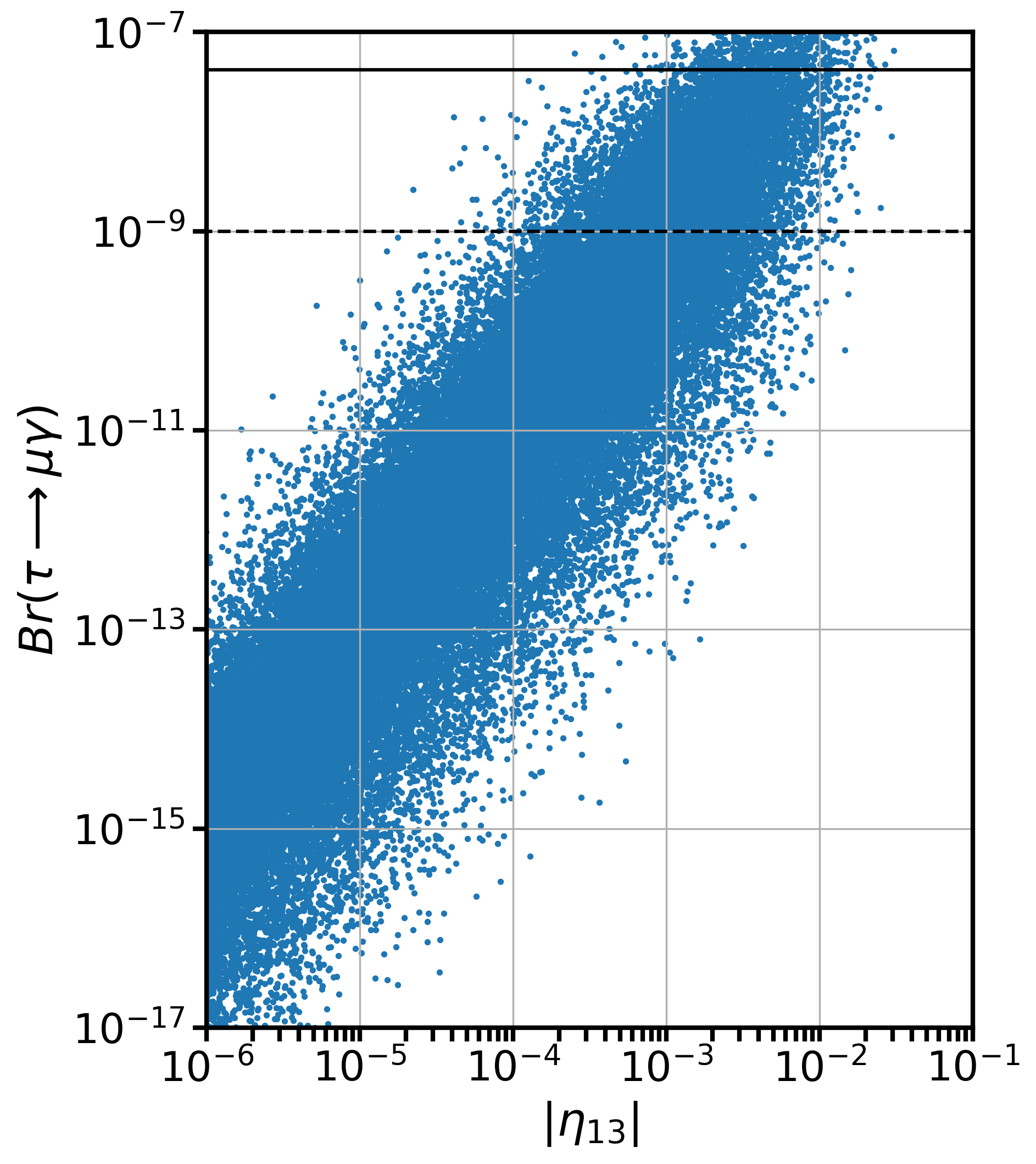}
    \end{subfigure}
\begin{subfigure}{0.3\textwidth}
    \includegraphics[width=\textwidth]{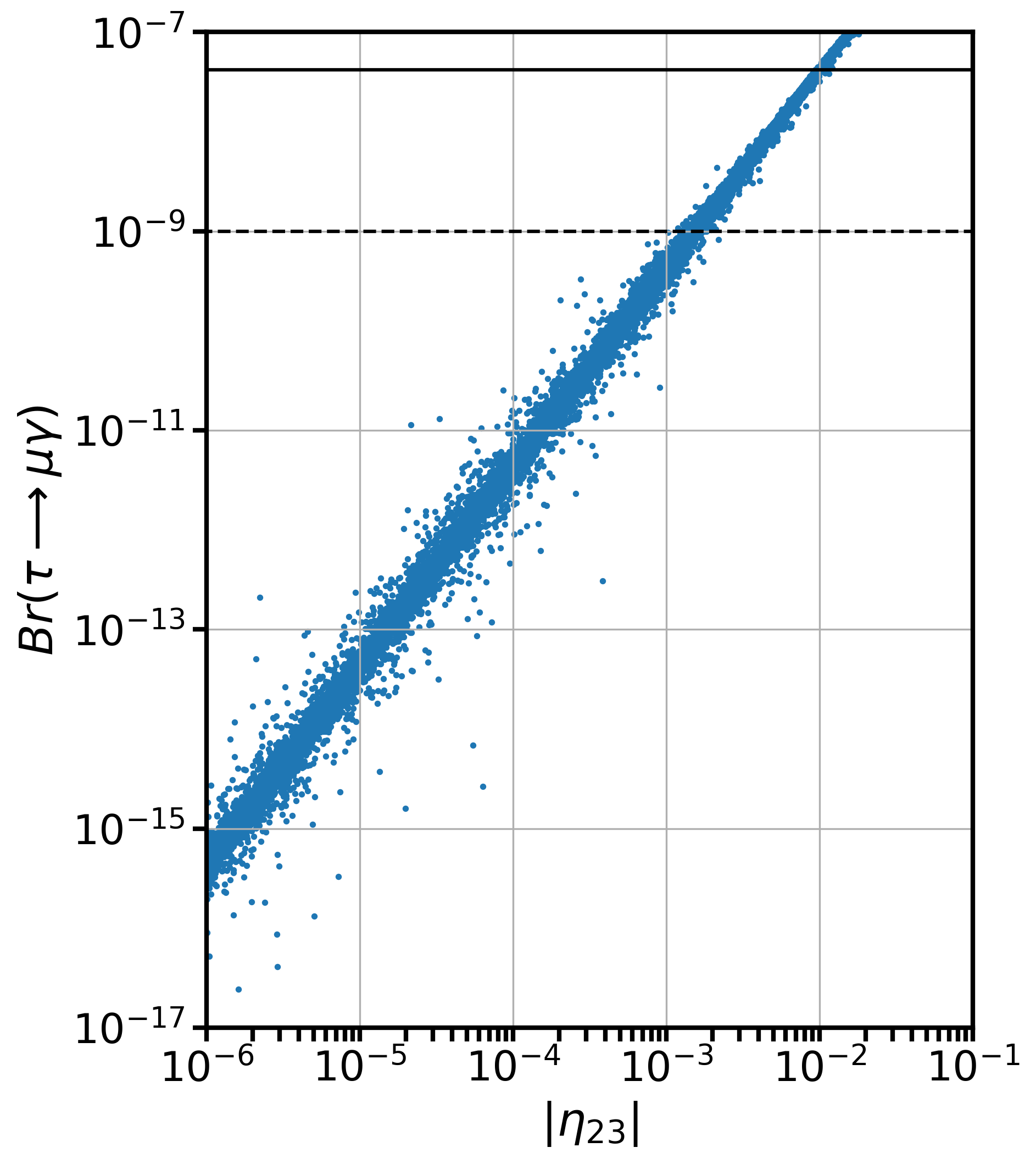}
    \end{subfigure}
      
\caption{\label{tau_to_muonIO} Scan of the branching ratio for the processes $\ell_{i} \longrightarrow \ell_{j}\gamma$ versus the nonunitary parameters in the IO case. The solid line accounts for the current constraints, while the dashed ones represent the future expected sensitivity for these decays. }
\label{fig:tau_mu_inverse}
\end{figure}

\begin{figure}
\begin{subfigure}{0.3\textwidth}
    \includegraphics[width=\textwidth]{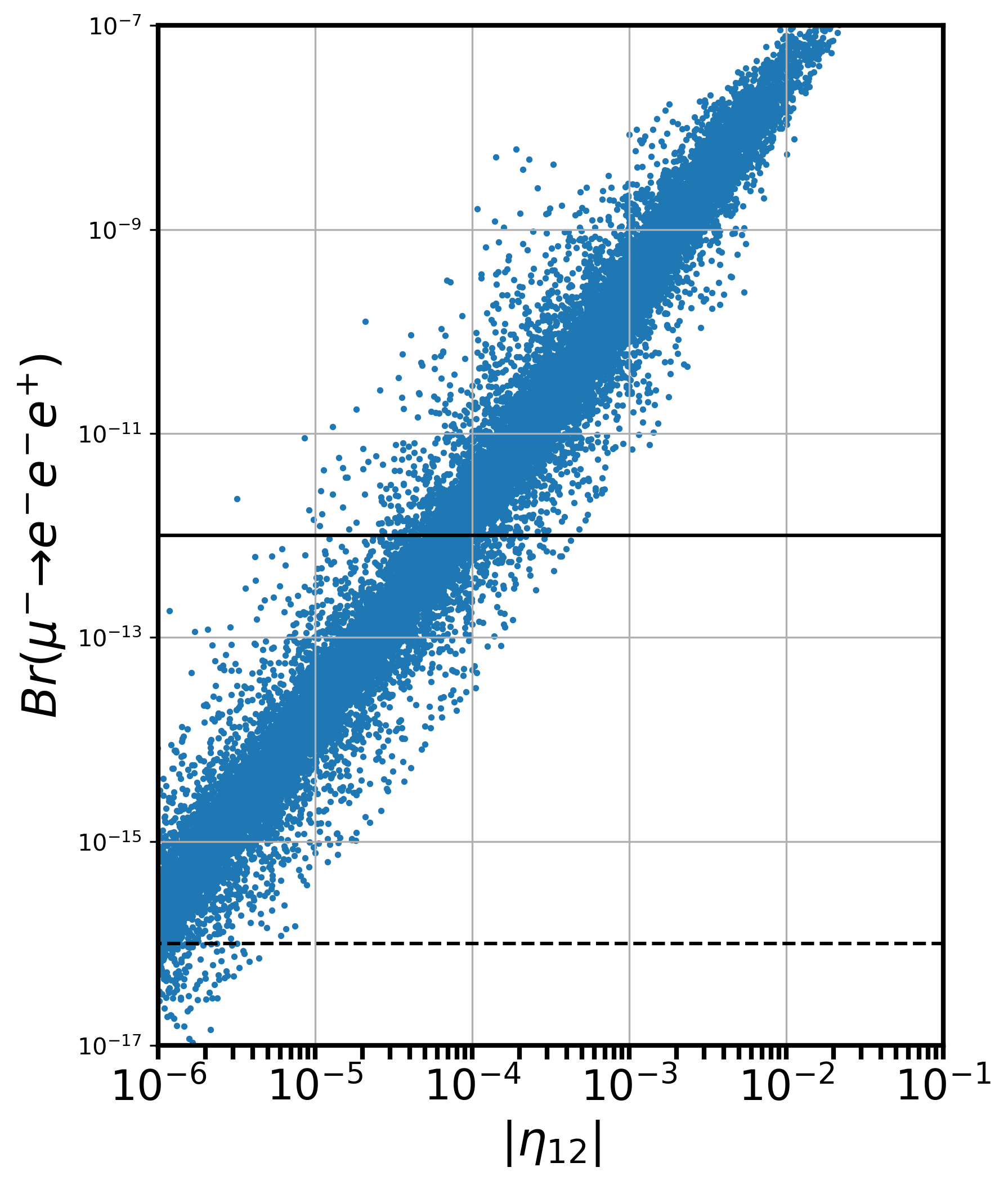}
    \end{subfigure}
    \begin{subfigure}{0.3\textwidth}
    \includegraphics[width=\textwidth]{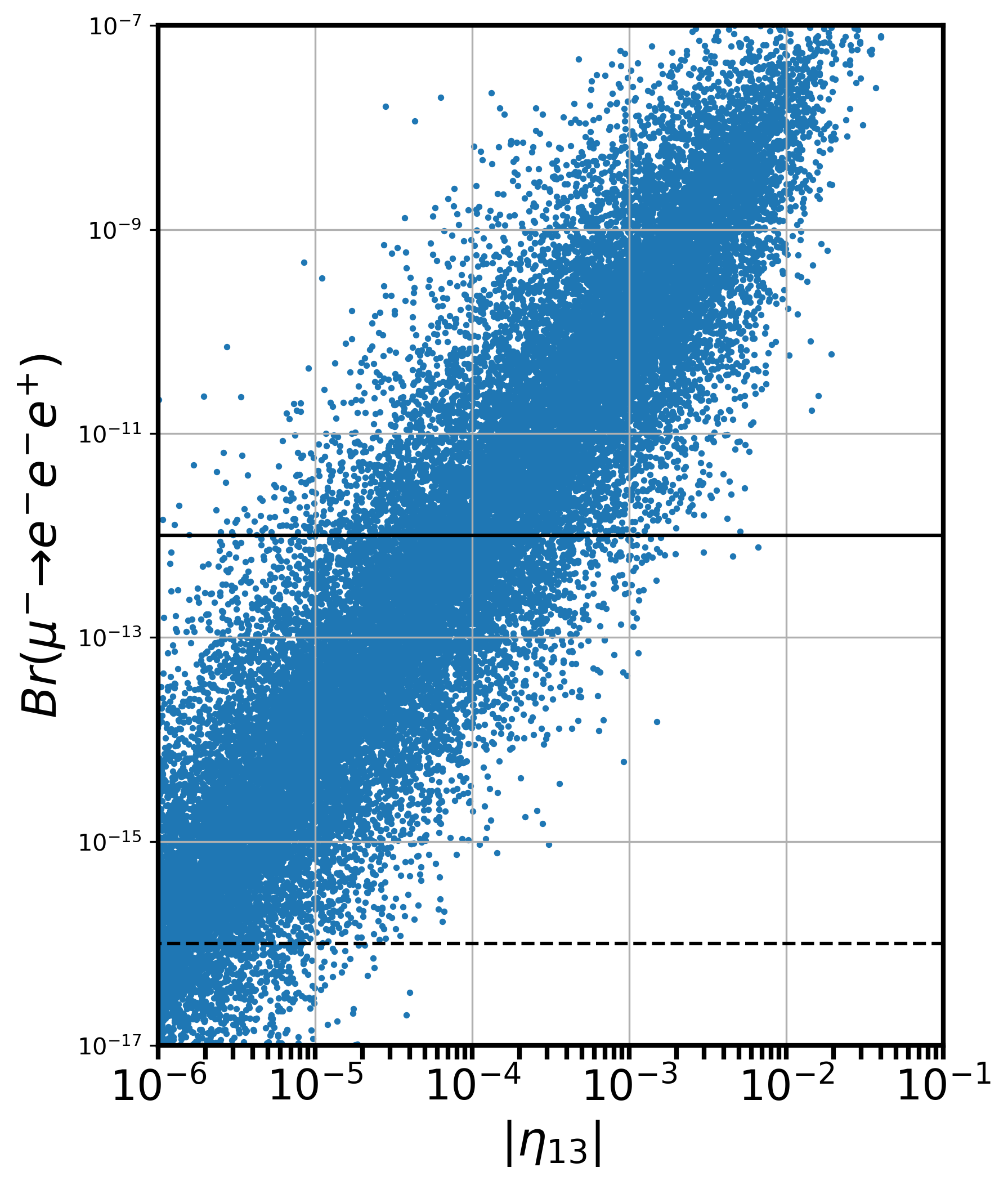}
    \end{subfigure}
\begin{subfigure}{0.3\textwidth}
    \includegraphics[width=\textwidth]{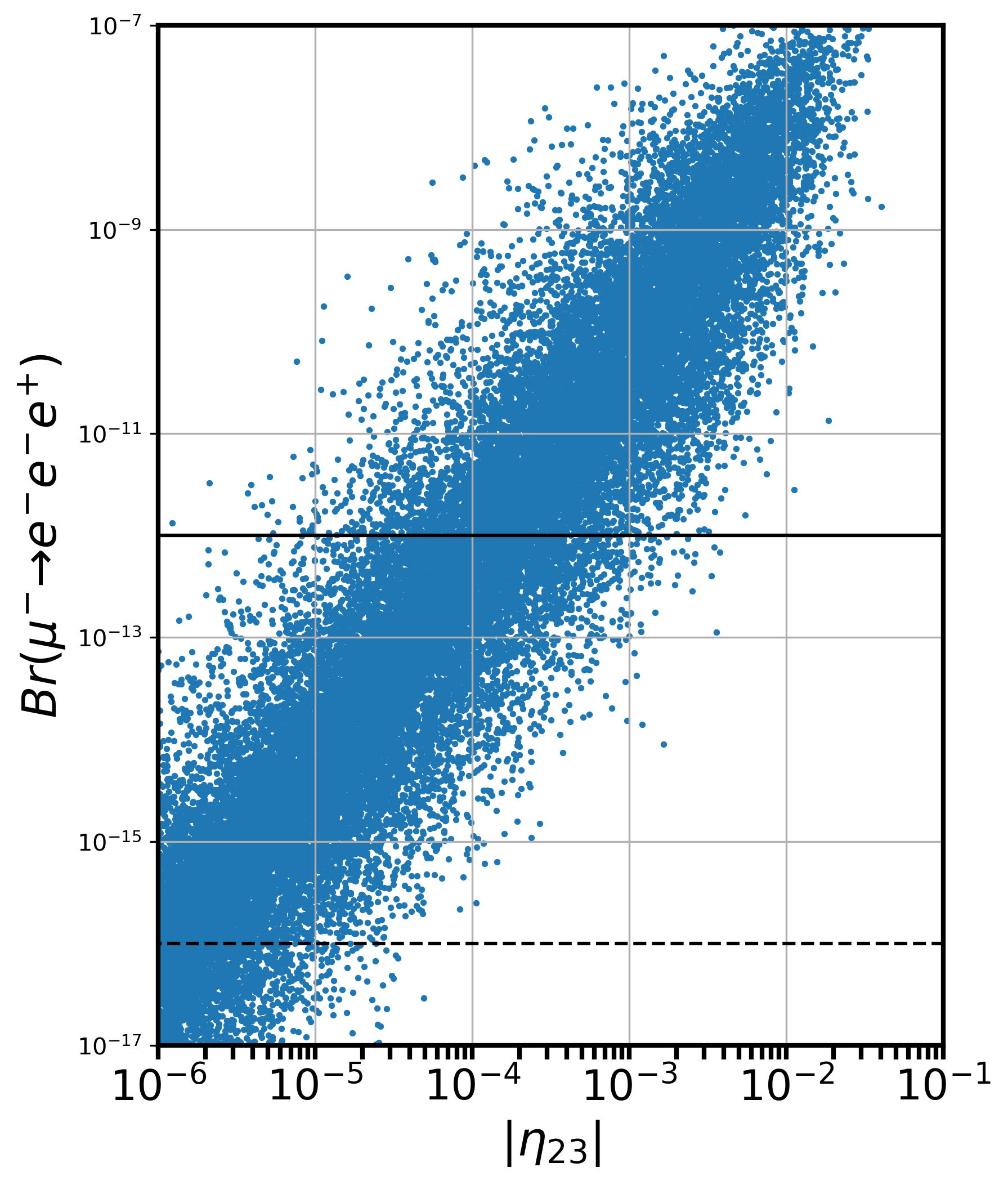}
    \end{subfigure} 
\caption{\label{mu_to_eee} Scan of the branching ratio for the processes $\mu^{-} \longrightarrow e^{-}e^{-}e^{+}$ versus the nonunitary parameters in the NO. The solid line accounts for the current constraints, while the dashed ones represent the future expected sensitivity for this decay. }
\label{fig:mueee}
\end{figure}
We perform a Monte Carlo analysis to find the current and future
sensitivity to non-unitary parameters from cLFV processes. We showed
our results in Figs.(~\ref{fig:figure1}-\ref{tau_to_muonIO}), where we
plotted the branching ratios of the cLFV processes versus the relevant
parameters for the linear seesaw scheme. In Fig.~\ref{fig:figure1}, we
ilustrate the constraints on $v_L$ using the cLFV
process. We can see from this plot that the $v_L$
  parameter, that gives us the scale of the $M_L$ matrix, should have
  a value above $10$~eV. On other hand, we have also computed the
Monte Carlo analysis for the cLFV in terms of the non-unitary
parameters. In particular, we have computed the branching ratios for
the $\mu\to e\gamma$, $\tau\to e\gamma$, and $\tau\to \mu\gamma$
depending on the non-unitary parameters $|\eta_{12}|$, $|\eta_{13}|$,
and $|\eta_{23}|$. The results, for normal ordering (NO), are shown in
Fig.~(\ref{muon_to_electron}), where we show the current constraint
for each process (Refs.~\cite{MEG:2013oxv,Belle-II:2018jsg}) with a
solid horizontal line and the future expected
constraint~\cite{MEG:2013oxv,BaBar:2009hkt} with a dashed
line. We can see from these plots that the $\mu \to
  e\gamma$ process is the one that most strongly reduces the possible
  values for the non-unitary parameters in comparison with the other
  two decays. 
For
the Inverted Ordering case (IO), we show the equivalent results in
Fig.~(\ref{tau_to_muonIO}). For this IO we also find that the $\mu \to e\gamma$
results are the most effective in reducing the possible non-unitary parameter values. From the future expected sensitivity, we can notice that it is expected that this tendency will continue in the future. 
We can
notice that the results are similar for both normal and inverted mass
ordering, except for the $|\eta_{23}|$ parameter, where the inverted ordering case suggest a more limited region of possible values for this parameter. Also, we can see the results from the $Br(u^{-}\to e^{-}e^{-}e^{+})$ in Fig. (\ref{mu_to_eee}). As in the case of $\mu \to e \gamma$, we see a correlation between the $\mu^{-}\to e^{-}e^{-}e^{+} $ and $|\eta_{12}|$. We can remark that in the case of the Mu3e experiment, the figure illustrates that a future result from this experiment could be sensitive to the non-unitary parameter $\alpha_{12}$ approximately at the order of $10^{-6}$.

Using Eq.~(\ref{matching}),
we notice that the same analysis can be done in terms of the $\alpha$
parametrization. It is easy to notice that the results will be
similar, in the sense that the suggested regions for $\alpha$ will be
of the same order as the $\eta$ parametrization. By comparing with previous works in the literature, for instance Ref.~\cite{Fernandez-Martinez:2016lgt} where a more general analysis on heavy neutral leptons was done, 
we can notice that while most of the parameters in our analysis tends to be of the same order as in this reference, the $\alpha_{13}$ parameter in our linear seesaw case appears to be allowed only at one order of magnitude below the reported constraint of $1.4\times10^{-3}$~\cite{Fernandez-Martinez:2016lgt}.

In summary, we studied non-unitary signatures of the Linear Seesaw model in charged lepton flavor violation processes.
For this analysis, we used the current constraints on the oscillation parameters (for normal and inverted ordering) and on the lightest mass $m_1$ given by cosmological constraints. 
We use the Casas-Ibarra parametrization to compute numerically the $M_D$ entrances in terms of the $M$ and $M_L$ matrix values. In order to make our results comparable with analysis from different observables, we wrote explicitly the relation between the so-called $\eta$ and $\alpha$ parametrizations that came from Eq.~(\ref{matching}) ($\eta_{ij}=\frac{1}{2}\alpha_{ij}$).
The current constraints for the $\alpha$ parameters~\cite{Forero:2021azc} can be translated into those for the $\eta$ parameters to have an idea of the sensitivity of cLFV processes in the Linear seesaw case. We found that the $\mu \longrightarrow e \gamma$ process could be more sensitive than these current oscillation experimental constraints.

\section{Acknowledgements}

We thank Gerardo Hern\'andez-Tom\'e and Eduardo Peinado for useful discussions. This work has been partially supported by CONAHCyT research grant:
A1-S-23238. The work of O. G. M. has also been supported by
SNII-Mexico.


\begin{thebibliography}{}

\end{thebibliography}


\begin{thebibliography}{10}


\bibitem{Minkowski:1977sc}
Peter Minkowski.
\newblock {$\mu \to e\gamma$ at a Rate of One Out of $10^{9}$ Muon Decays?}
\newblock {\em Phys. Lett. B}, 67:421--428, 1977.


\bibitem{Mohapatra:1979ia}
Rabindra~N. Mohapatra and Goran Senjanovic.
\newblock {Neutrino Mass and Spontaneous Parity Nonconservation}.
\newblock {\em Phys. Rev. Lett.}, 44:912, 1980.


\bibitem{Gell-Mann:1979vob}
Murray Gell-Mann, Pierre Ramond, and Richard Slansky.
\newblock {Complex Spinors and Unified Theories}.
\newblock {\em Conf. Proc. C}, 790927:315--321, 1979.


\bibitem{Yanagida:1979as}
Tsutomu Yanagida.
\newblock {Horizontal gauge symmetry and masses of neutrinos}.
\newblock {\em Conf. Proc. C}, 7902131:95--99, 1979.


\bibitem{Schechter:1980gr}
J.~Schechter and J.~W.~F. Valle.
\newblock {Neutrino Masses in SU(2) x U(1) Theories}.
\newblock {\em Phys. Rev. D}, 22:2227, 1980.


\bibitem{Foot:1988aq}
Robert Foot, H.~Lew, X.~G. He, and Girish~C. Joshi.
\newblock {Seesaw Neutrino Masses Induced by a Triplet of Leptons}.
\newblock {\em Z. Phys. C}, 44:441, 1989.


\bibitem{Mohapatra:1986bd}
R.~N. Mohapatra and J.~W.~F. Valle.
\newblock {Neutrino Mass and Baryon Number Nonconservation in Superstring
  Models}.
\newblock {\em Phys. Rev. D}, 34:1642, 1986.


\bibitem{Akhmedov:1995vm}
Evgeny~K. Akhmedov, Manfred Lindner, Erhard Schnapka, and J.~W.~F. Valle.
\newblock {Dynamical left-right symmetry breaking}.
\newblock {\em Phys. Rev. D}, 53:2752--2780, 1996.


\bibitem{Malinsky:2005bi}
Michal Malinsky, J.~C. Romao, and J.~W.~F. Valle.
\newblock {Novel supersymmetric SO(10) seesaw mechanism}.
\newblock {\em Phys. Rev. Lett.}, 95:161801, 2005.


\bibitem{Batra:2023mds}
Aditya Batra, Praveen Bharadwaj, Sanjoy Mandal, Rahul Srivastava, and Jos\'e
  W.~F. Valle.
\newblock {Phenomenology of the simplest linear seesaw mechanism}.
\newblock {\em JHEP}, 07:221, 2023.


\bibitem{Miranda:2020syh}
O.~G. Miranda, D.~K. Papoulias, O.~Sanders, M.~T\'ortola, and J.~W.~F. Valle.
\newblock {Future CEvNS experiments as probes of lepton unitarity and
  light-sterile neutrinos}.
\newblock {\em Phys. Rev. D}, 102:113014, 2020.


\bibitem{Escrihuela:2015wra}
F.~J. Escrihuela, D.~V. Forero, O.~G. Miranda, M.~Tortola, and J.~W.~F. Valle.
\newblock {On the description of nonunitary neutrino mixing}.
\newblock {\em Phys. Rev. D}, 92(5):053009, 2015.
\newblock [Erratum: Phys.Rev.D 93, 119905 (2016)].


\bibitem{CentellesChulia:2024uzv}
Salvador Centelles~Chuli\'a, Antonio Herrero-Brocal, and Avelino Vicente.
\newblock {The Type-I Seesaw family}.
\newblock {\em JHEP}, 07:060, 2024.


\bibitem{Garnica:2023ccx}
J.~C. Garnica, G.~Hern\'andez-Tom\'e, and E.~Peinado.
\newblock {Charged lepton-flavor violating processes and suppression of
  nonunitary mixing effects in low-scale seesaw models}.
\newblock {\em Phys. Rev. D}, 108(3):035033, 2023.


\bibitem{Forero:2011pc}
D.~V. Forero, S.~Morisi, M.~Tortola, and J.~W.~F. Valle.
\newblock {Lepton flavor violation and non-unitary lepton mixing in low-scale
  type-I seesaw}.
\newblock {\em JHEP}, 09:142, 2011.


\bibitem{Kanaya:1980cw}
Kazuyuki Kanaya.
\newblock {Neutrino Mixing in the Minimal SO(10) Model}.
\newblock {\em Prog. Theor. Phys.}, 64:2278, 1980.


\bibitem{Schechter:1981cv}
J.~Schechter and J.~W.~F. Valle.
\newblock {Neutrino Decay and Spontaneous Violation of Lepton Number}.
\newblock {\em Phys. Rev. D}, 25:774, 1982.


\bibitem{Gronau:1984ct}
Michael Gronau, Chung~Ngoc Leung, and Jonathan~L. Rosner.
\newblock {Extending Limits on Neutral Heavy Leptons}.
\newblock {\em Phys. Rev. D}, 29:2539, 1984.


\bibitem{Nardi:1994iv}
Enrico Nardi, Esteban Roulet, and Daniele Tommasini.
\newblock {Limits on neutrino mixing with new heavy particles}.
\newblock {\em Phys. Lett. B}, 327:319--326, 1994.


\bibitem{Atre:2009rg}
Anupama Atre, Tao Han, Silvia Pascoli, and Bin Zhang.
\newblock {The Search for Heavy Majorana Neutrinos}.
\newblock {\em JHEP}, 05:030, 2009.


\bibitem{Escrihuela:2016ube}
F.~J. Escrihuela, D.~V. Forero, O.~G. Miranda, M.~T\'ortola, and J.~W.~F.
  Valle.
\newblock {Probing CP violation with non-unitary mixing in long-baseline
  neutrino oscillation experiments: DUNE as a case study}.
\newblock {\em New J. Phys.}, 19(9):093005, 2017.


\bibitem{Fernandez-Martinez:2016lgt}
Enrique Fernandez-Martinez, Josu Hernandez-Garcia, and Jacobo Lopez-Pavon.
\newblock {Global constraints on heavy neutrino mixing}.
\newblock {\em JHEP}, 08:033, 2016.


\bibitem{Blennow:2023mqx}
Mattias Blennow, Enrique Fern\'andez-Mart\'\i{}nez, Josu
  Hern\'andez-Garc\'\i{}a, Jacobo L\'opez-Pav\'on, Xabier Marcano, and Daniel
  Naredo-Tuero.
\newblock {Bounds on lepton non-unitarity and heavy neutrino mixing}.
\newblock {\em JHEP}, 08:030, 2023.


\bibitem{Forero:2021azc}
D.~V. Forero, C.~Giunti, C.~A. Ternes, and M.~Tortola.
\newblock {Nonunitary neutrino mixing in short and long-baseline experiments}.
\newblock {\em Phys. Rev. D}, 104(7):075030, 2021.


\bibitem{Denton_2022}
Peter~B. Denton and Julia Gehrlein.
\newblock New oscillation and scattering constraints on the tau row matrix
  elements without assuming unitarity.
\newblock {\em Journal of High Energy Physics}, 2022(6), jun 2022.


\bibitem{Dutta:2019hmb}
Debajyoti Dutta and Samiran Roy.
\newblock {Non-Unitarity at DUNE and T2HK with Charged and Neutral Current
  Measurements}.
\newblock {\em J. Phys. G}, 48(4):045004, 2021.


\bibitem{Giunti:2007ry}
Carlo Giunti and Chung~W. Kim.
\newblock {\em {Fundamentals of Neutrino Physics and Astrophysics}}.
\newblock 2007.


\bibitem{Rodejohann:2011vc}
W.~Rodejohann and J.~W.~F. Valle.
\newblock {Symmetrical Parametrizations of the Lepton Mixing Matrix}.
\newblock {\em Phys. Rev. D}, 84:073011, 2011.


\bibitem{Valle:2015pba}
Jose W.~F. Valle and Jorge~C. Romao.
\newblock {\em {Neutrinos in high energy and astroparticle physics}}.
\newblock Physics textbook. Wiley-VCH, Weinheim, 2015.


\bibitem{Chatterjee:2021xyu}
Sabya~Sachi Chatterjee, O.~G. Miranda, M.~T\'ortola, and J.~W.~F. Valle.
\newblock {Nonunitarity of the lepton mixing matrix at the European Spallation
  Source}.
\newblock {\em Phys. Rev. D}, 106(7):075016, 2022.


\bibitem{Celestino-Ramirez:2023zox}
Jes\'us~Miguel Celestino-Ram\'\i{}rez, F.~J. Escrihuela, L.~J. Flores, and
  O.~G. Miranda.
\newblock {Testing the nonunitarity of the leptonic mixing matrix at FASERv and
  FASERv2}.
\newblock {\em Phys. Rev. D}, 109(1):L011705, 2024.


\bibitem{CentellesChulia:2024sff}
Salvador Centelles~Chuli\'a, O.~G. Miranda, and Jose W.~F. Valle.
\newblock {Leptonic neutral-current probes in a short-distance DUNE-like
  setup}.
\newblock {\em Phys. Rev. D}, 109(11):115007, 2024.


\bibitem{Xing:2007zj}
Zhi-zhong Xing.
\newblock {Correlation between the Charged Current Interactions of Light and
  Heavy Majorana Neutrinos}.
\newblock {\em Phys. Lett. B}, 660:515--521, 2008.


\bibitem{Xing:2011ur}
Zhi-zhong Xing.
\newblock {A full parametrization of the 6 X 6 flavor mixing matrix in the
  presence of three light or heavy sterile neutrinos}.
\newblock {\em Phys. Rev. D}, 85:013008, 2012.


\bibitem{Han:2021qum}
He-chong Han and Zhi-zhong Xing.
\newblock {A full parametrization of the 9 x 9 active-sterile flavor mixing
  matrix in the inverse or linear seesaw scenario of massive neutrinos}.
\newblock {\em Nucl. Phys. B}, 973:115609, 2021.


\bibitem{Blennow:2016jkn}
Mattias Blennow, Pilar Coloma, Enrique Fernandez-Martinez, Josu
  Hernandez-Garcia, and Jacobo Lopez-Pavon.
\newblock {Non-Unitarity, sterile neutrinos, and Non-Standard neutrino
  Interactions}.
\newblock {\em JHEP}, 04:153, 2017.

\bibitem{MEG:2013oxv}
J.~Adam et~al.
\newblock {New constraint on the existence of the $\mu^+ \to e^+\gamma$ decay}.
\newblock {\em Phys. Rev. Lett.}, 110:201801, 2013.


\bibitem{MEGII:2018kmf}
A.~M. Baldini et~al.
\newblock {The design of the MEG II experiment}.
\newblock {\em Eur. Phys. J. C}, 78(5):380, 2018.


\bibitem{BaBar:2009hkt}
Bernard Aubert et~al.
\newblock {Searches for Lepton Flavor Violation in the Decays tau+-
  ---\ensuremath{>} e+- gamma and tau+- ---\ensuremath{>} mu+- gamma}.
\newblock {\em Phys. Rev. Lett.}, 104:021802, 2010.


\bibitem{Belle-II:2018jsg}
W.~Altmannshofer et~al.
\newblock {The Belle II Physics Book}.
\newblock {\em PTEP}, 2019(12):123C01, 2019.
\newblock [Erratum: PTEP 2020, 029201 (2020)].

\bibitem{SINDRUM:1987nra}
Bellgardt, U. and others.
\newblock{Search for the Decay $\mu^+ \to e^+ e^+ e^-$}.
\newblock{\em Nucl. Phys. B}, 299, 1--6, 1988.

\bibitem{Hesketh:2022wgw}
G.~Hesketh \textit{et al.} [Mu3e],
[arXiv:2204.00001 [hep-ex]].

\bibitem{Mu3e:2020gyw}
K.~Arndt \textit{et al.} [Mu3e],
Nucl. Instrum. Meth. A \textbf{1014} (2021), 165679
doi:10.1016/j.nima.2021.165679
[arXiv:2009.11690 [physics.ins-det]].

\bibitem{He:2002pva}
Bo~He, T.~P. Cheng, and Ling-Fong Li.
\newblock {A Less suppressed mu ---\ensuremath{>} e gamma loop amplitude and
  extra dimension theories}.
\newblock {\em Phys. Lett. B}, 553:277--283, 2003.

\bibitem{Ilakovac:1994kj}
A.~Ilakovac and A.~Pilaftsis.
\newblock {Flavor violating charged lepton decays in seesaw-type models}.
\newblock {\em Nucl. Phys. B}, 437:491, 1995.


\bibitem{deSalas:2020pgw}
P.~F. de~Salas, D.~V. Forero, S.~Gariazzo, P.~Mart\'\i{}nez-Mirav\'e, O.~Mena,
  C.~A. Ternes, M.~T\'ortola, and J.~W.~F. Valle.
\newblock {2020 global reassessment of the neutrino oscillation picture}.
\newblock {\em JHEP}, 02:071, 2021.

\bibitem{DAmbrosio:2002vsn}
G.~D'Ambrosio, G.~F. Giudice, G.~Isidori, and A.~Strumia.
\newblock {Minimal flavor violation: An Effective field theory approach}.
\newblock {\em Nucl. Phys. B}, 645:155--187, 2002.


\bibitem{Casas:2001sr}
J.~A. Casas and A.~Ibarra.
\newblock {Oscillating neutrinos and $\mu \to e, \gamma$}.
\newblock {\em Nucl. Phys. B}, 618:171--204, 2001.






\end{thebibliography}

\end{document}